\documentclass[a4paper,12pt]{article}
\usepackage{amsfonts}
\usepackage{graphicx}
\usepackage{subcaption}
\usepackage{pstricks}
\usepackage{pstricks-add}
\usepackage{tikz}
\usepackage{verbatim}
\usepackage{dsfont}
\usetikzlibrary{decorations.pathreplacing}
\usetikzlibrary{decorations.markings}
\usepackage{amsmath}
\usepackage{amssymb}
\usepackage[title]{appendix}
\usepackage[margin=2.3cm]{geometry}

\DeclareMathOperator{\Tr}{Tr}

\newcommand{\beq}{\begin{equation}}
\newcommand{\dtt}{$D_2^2$ }
\newcommand{\eeq}{\end{equation}}
\newcommand{\no}{\noindent}

\begin{document}

\title{Integrable boundary conditions \\ in the antiferromagnetic Potts model}
\date{}

\maketitle

\begin{center}

{ Niall F. Robertson$^{1}$, Michal Pawelkiewicz$^{1}$, Jesper Lykke Jacobsen$^{1,2,3}$, and Hubert Saleur$^{1,4}$}

\vspace{1.0cm}
{\sl\small $^1$ Universit\'e Paris Saclay, CNRS, CEA, Institut de Physique Th\'eorique, \\ F-91191 Gif-sur-Yvette, France\\}
{\sl\small $^2$ Laboratoire de Physique de l'\'Ecole Normale Sup\'erieure, ENS, Universit\'e PSL, \\
CNRS, Sorbonne Universit\'e, Universit\'e de Paris, F-75005 Paris, France\\}
{\sl\small $^3$ Sorbonne Universit\'e, \'Ecole Normale Sup\'erieure, CNRS, \\
Laboratoire de Physique (LPENS), F-75005 Paris, France\\}
{\sl\small $^4$ Department of Physics and Astronomy,
University of Southern California, \\
Los Angeles, CA 90089, USA\\}

\end{center}

\vskip 2cm

\begin{abstract}
We present an exact mapping between the staggered six-vertex model and an integrable model constructed from the twisted affine \dtt Lie algebra. Using the known relations between the staggered six-vertex model and the antiferromagnetic Potts model, this mapping allows us to study the latter model using tools from integrability.  We show that there is a simple interpretation of one of the known $K$-matrices of the \dtt model in terms of Temperley-Lieb algebra generators, and use this to present an integrable Hamiltonian that turns out to be in the same universality class as the antiferromagnetic Potts model with free boundary conditions. 
The intriguing degeneracies in the spectrum observed in related works (\cite{martinsd22} and \cite{nepomechied22} ) are discussed.  
\end{abstract}

\tableofcontents

\section{Introduction}\label{intro}
The critical antiferromagnetic Potts model has been the subject of intense study for many years \cite{B-AF,AFPottsSaleur,JACOBSEN2006}. A striking feature is that the conformal field theory describing its continuum limit is ``non-compact'', leading to the observation of a continuum of critical exponents \cite{ikhlef2008,Ikhlef2012}.
Due to this unusual feature, this model has subsequently become the subject of many pieces of work \cite{Candu2013,Frahm2014,Bazhanov2019}.
Interestingly, the very same continuum limit is shared by a model of polymer collapse, driven to the so-called theta-point by a critial attraction between monomers \cite{Vernier2014,Vernier2015}.

\smallskip

Recent work \cite{Robertson2019} on the critical antiferromagnetic (AF) Potts model has identified new conformally invariant boundary conditions. The work in \cite{Robertson2019} used numerical methods to study the corresponding boundary conformal field theory describing the antiferromagnetic Potts model, but an exact solution of the open model, even in the simplest case of free boundary conditions, was not considered. The purpose of this paper is to extend the work of \cite{Robertson2019} by applying the tools of integrability. To our knowledge, the transfer matrix describing free boundary conditions in the AF Potts model, first studied in \cite{AFPottsSaleur}, is not solvable by Bethe Ansatz. However, here we use the Bethe Ansatz to study a particular boundary condition found in the context of the integrable \dtt model, and we show that this boundary condition is in the same universality class as the free AF Potts model.

\smallskip

The paper is structured as follows: in section \ref{s6vd22} the formulation of the antiferromagnetic Potts model as a staggered six-vertex model is reviewed. It is shown that there is an exact mapping between the staggered six-vertex model and the integrable model constructed from the twisted affine \dtt Lie algebra. In section \ref{secopengen} the model with open boundary conditions is considered. A particular $K$-matrix from the \dtt model \cite{martinsd22,nepomechied22} is interpreted in the context of the staggered six-vertex model. In particular, it is found that the Hamiltonian of the model with the boundary conditions described by this $K$-matrix has a very simple interpretation in terms of generators of the Temperley-Lieb algebra. This integrable, open Hamiltonian is written in equation (\ref{hopen4}). The symmetry group of the chain is discussed and the additional degeneracies of the \dtt chain that were observed in \cite{martinsd22} and \cite{nepomechied22} are interpreted using  a symmetry operator written in terms of Temperley-Lieb algebra generators. In section \ref{d22type2} the Bethe Ansatz solution of the model with these boundary conditions is presented, and the critical exponents are derived analytically. Some numerical solutions to the Bethe Ansatz equations are presented and are used to show that the scaling behaviour ot the chain is the same as that of the antiferromagnetic Potts model with free boundary conditions. Section \ref{TLreps} considers the model in two different representations of the Temperley-Lieb algebra and numerical results confirms that we have indeed correctly identified the underlying boundary CFT.

\medskip

For the reader's convenience, we here give a list of notations, consistent with our earlier works on related topics: 
\begin{itemize}
\item$\mathcal{W}_j$ --- standard modules over $\mathsf{TL}_{N}$,
\item$j$ --- the $U_q(sl(2))$ spin, with $l=2j$ the number of through-lines,
\item$L$ --- number of Potts spins in a horizontal row of the lattice,
\item$N$ --- number of strands in the loop model, or the number of spin-$\frac{1}{2}$ sites in the spin chain,
\item$c^m_l$ --- string function, i.e., the generating function of levels in the $Z_{k-2}$ parafermion CFT.
\end{itemize}
We draw particular attention to the distinction between $L$ and $N\equiv 2L$. The Potts model with $L$ Potts spins in the horizontal direction is described by a spin chain with $N=2L$ spin-$\frac{1}{2}$ sites. The \dtt vertex model of width $L$ also corresponds to a spin chain with $N=2L$ spin-$\frac{1}{2}$ sites.

\section{The staggered six-vertex model and the \dtt model}\label{s6vd22}

\subsection{Background}
The two-dimensional $Q$-state Potts model is defined by the classical Hamiltonian

\beq\label{classicalpotts}
\mathcal{H} = - K\sum\limits_{\langle ij \rangle }\delta_{\sigma_i \sigma_j} \,,
\eeq
where $\sigma_i = 1,2,\ldots,Q$ and $\langle ij \rangle$ denotes the set of nearest neighbours on the square lattice. This model
has been reviewed in many places \cite{Robertson2019, AFPottsSaleur, ikhlef2008}. It is well known that the Potts model can be reformulated as a height, loop and vertex model \cite{ikhlef2010} where the partition functions are identical to that of the original Potts model described in terms of spins, but with different observables. It is another well-known result that when the correspondence between the Potts and the vertex model is carried out at the so-called ``ferromagnetic critical point'', the resulting vertex model is the celebrated ``six-vertex model''. Carrying out this Potts/vertex mapping at the other critical point of the Potts model, the ``antiferromagnetic critical point'', one obtains the ``staggered six-vertex model'' where the Boltzmann weights take particular values that alternate with each row/column.

\smallskip

Here we will show that the staggered six-vertex model is identical to an integrable model constructed from the \dtt affine Lie algebra. This relationship between the \dtt model and the staggered six vertex model was first alluded to in \cite{frahmsd22} where the spectra of the two models were shown to be identical. Here we take this result further and show that the transfer matrices of the two models can in fact be identified. This paves the way in later sections to derive new results related to the antiferromagnetic Potts model and its integrable boundary conditions. We will be particularly interested in ``free'' boundary conditions in the Potts model which corresponds in (\ref{classicalpotts}) to imposing no additional constraint on the Potts spins at the boundary so that the sum runs over all nearest neighbours as usual but boundary spins have fewer nearest neighbours.


\subsection{Review of the staggered six-vertex model}

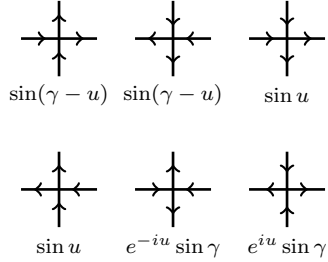
\begin{figure}
	\centering
		\begin{tikzpicture}[scale=1]
		
		\node at (0.5,0.25) {\scriptsize{$\sin(\gamma-u)$}};
		\node at (2,0.25) {\scriptsize{$\sin(\gamma-u)$}};
		\node at (3.5,0.25) {\scriptsize{$\sin u$}};
		\node at (0.5,-1.75) {\scriptsize{$\sin u$}};
		\node at (2,-1.75) {\scriptsize{$e^{-iu}\sin \gamma$}};
		\node at (3.5,-1.75) {\scriptsize{$e^{iu}\sin \gamma$}};

			\begin{scope}[thick,decoration={
			    markings,
			    mark=at position 0.65 with {\arrow{>}}}
			    ] 
			\draw[black,line width = 1pt,postaction={decorate}] (0,1)--(0.5,1);
			\draw[black,line width = 1pt,postaction={decorate}] (0.5,0.5)--(0.5,1);
			\draw[black,line width = 1pt,postaction={decorate}] (0.5,1)--(1,1);
			 \draw[black,line width = 1pt,postaction={decorate}](0.5,1)--(0.5,1.5);
			\end{scope}

			\begin{scope}[thick,decoration={
			    markings,
			    mark=at position 0.55 with {\arrow{<}}}
			    ] 
			\draw[black,line width = 1pt,postaction={decorate}] (1.5,1)--(2,1);
			\draw[black,line width = 1pt,postaction={decorate}] (2,0.5)--(2,1);
			\draw[black,line width = 1pt,postaction={decorate}] (2,1)--(2.5,1);
			 \draw[black,line width = 1pt,postaction={decorate}](2,1)--(2,1.5);
			\end{scope}

			\begin{scope}[thick,decoration={
			    markings,
			    mark=at position 0.65 with {\arrow{>}}}
			    ] 
			\draw[black,line width = 1pt,postaction={decorate}] (3,1)--(3.5,1);
			\draw[black,line width = 1pt,postaction={decorate}] (3.5,1)--(3.5,0.5);
			\draw[black,line width = 1pt,postaction={decorate}] (3.5,1)--(4,1);
			 \draw[black,line width = 1pt,postaction={decorate}](3.5,1.5)--(3.5,1);
			\end{scope}

			\begin{scope}[thick,decoration={
			    markings,
			    mark=at position 0.55 with {\arrow{<}}}
			    ] 
			\draw[black,line width = 1pt,postaction={decorate}] (0,-1)--(0.5,-1);
			\draw[black,line width = 1pt,postaction={decorate}] (0.5,-1)--(0.5,-1.5);
			\draw[black,line width = 1pt,postaction={decorate}] (0.5,-1)--(1,-1);
			 \draw[black,line width = 1pt,postaction={decorate}](0.5,-0.5)--(0.5,-1);
			\end{scope}		

			\begin{scope}[thick,decoration={
 			   markings,
			   mark=at position 0.55 with {\arrow{<}}}
			   ] 
			   \draw[black,line width = 1pt,postaction={decorate}] (2,-1)--(1.5,-1);
			   \draw[black,line width = 1pt,postaction={decorate}] (2,-1.5)--(2,-1);
			   \draw[black,line width = 1pt,postaction={decorate}] (2,-1)--(2.5,-1);
			   \draw[black,line width = 1pt,postaction={decorate}](2,-0.5)--(2,-1);
			   \end{scope}
			   
   			\begin{scope}[thick,decoration={
    			   markings,
   			   mark=at position 0.55 with {\arrow{>}}}
   			   ] 
   			   \draw[black,line width = 1pt,postaction={decorate}] (3.5,-1)--(3,-1);
   			   \draw[black,line width = 1pt,postaction={decorate}] (3.5,-1.5)--(3.5,-1);
   			   \draw[black,line width = 1pt,postaction={decorate}] (3.5,-1)--(4,-1);
   			   \draw[black,line width = 1pt,postaction={decorate}](3.5,-0.5)--(3.5,-1);
   			   \end{scope}

			\end{tikzpicture}
			\caption{The six vertices and their Boltzmann weights.}\label{sixvertices}
			\end{figure}

\begin{figure}
	\centering
	\begin{tikzpicture}
		
\begin{scope}[thick,decoration={
    markings,
    mark=at position 0.55 with {\arrow{>}}}
    ] 
\draw[black,line width = 1pt,postaction={decorate}] (0,1)--(1,1);
\end{scope}	

\begin{scope}[thick,decoration={
    markings,
    mark=at position 0.55 with {\arrow{>}}}
    ] 
\draw[black,line width = 1pt,postaction={decorate}] (0,2)--(1,2);
\end{scope}	

\begin{scope}[thick,decoration={
    markings,
    mark=at position 0.55 with {\arrow{>}}}
    ] 
\draw[black,line width = 1pt,postaction={decorate}] (0,3)--(1,3);
\end{scope}	

\begin{scope}[thick,decoration={
    markings,
    mark=at position 0.55 with {\arrow{>}}}
    ] 
\draw[black,line width = 1pt,postaction={decorate}] (0,4)--(1,4);
\end{scope}	

\begin{scope}[thick,decoration={
    markings,
    mark=at position 0.55 with {\arrow{>}}}
    ] 
\draw[black,line width = 1pt,postaction={decorate}] (1,0)--(1,1);
\end{scope}	

\begin{scope}[thick,decoration={
    markings,
    mark=at position 0.55 with {\arrow{>}}}
    ] 
\draw[black,line width = 1pt,postaction={decorate}] (2,0)--(2,1);
\end{scope}	

\begin{scope}[thick,decoration={
    markings,
    mark=at position 0.55 with {\arrow{>}}}
    ] 
\draw[black,line width = 1pt,postaction={decorate}] (3,0)--(3,1);
\end{scope}	

\begin{scope}[thick,decoration={
    markings,
    mark=at position 0.55 with {\arrow{>}}}
    ] 
\draw[black,line width = 1pt,postaction={decorate}] (4,0)--(4,1);
\end{scope}	
		
\draw[black,line width = 1pt](1,1)--(5,1);
\draw[black,line width = 1pt](1,2)--(5,2);
\draw[black,line width = 1pt](1,3)--(5,3);
\draw[black,line width = 1pt](1,4)--(5,4);

\draw[black,line width = 1pt](1,1)--(1,5);
\draw[black,line width = 1pt](2,1)--(2,5);
\draw[black,line width = 1pt](3,1)--(3,5);
\draw[black,line width = 1pt](4,1)--(4,5);

\node at (-0.5,1) {$u$};
\node at (-0.5,2) {$u$};
\node at (-0.5,3) {$u$};
\node at (-0.5,4) {$u$};

\node at (1,-0.5) {$0$};
\node at (2,-0.5) {$0$};
\node at (3,-0.5) {$0$};
\node at (4,-0.5) {$0$};

\end{tikzpicture}
\caption{The spectral parameters on the lattice of the six-vertex model }\label{sixvertlattice}
\end{figure}
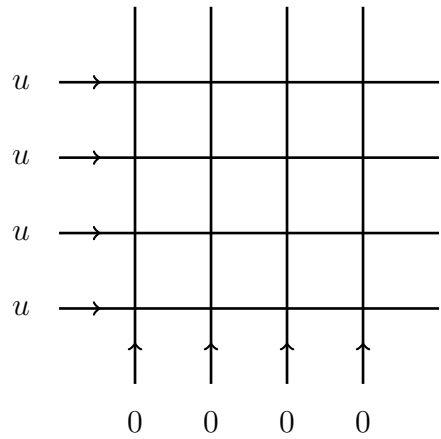

The six-vertex model with no staggering is defined by placing arrows on the edges of a square lattice subject to the constraint that there must be two incoming and two outgoing arrows at every vertex. The six possible vertices that satisfy this constraint are shown in figure \ref{sixvertices}. Each of these vertices then takes a particular Boltzmann weight parameterised by the `spectral parameter' $u$ which controls the amount of anisotropy. The Boltzmann weights are also functions of the crossing parameter $\gamma$ which appears in the $Q$-state Potts model as
\beq
 \sqrt{Q}=e^{i\gamma}+e^{-i\gamma} \,.
\eeq
We can encode the Boltzmann weights in the $\check{R}$-matrix which acts on the space
\beq
\{|\uparrow\uparrow\rangle, |\uparrow\downarrow\rangle, |\downarrow\uparrow\rangle, |\downarrow\downarrow\rangle \} \,.
\eeq
Taking
\beq\label{sixvertrmat}
\check{R}(u)=\begin{pmatrix}
	\sin(\gamma-u) & 0 & 0 & 0 \\
	0 & e^{-iu}\sin\gamma & \sin u & 0\\
	0 & \sin u & e^{iu}\sin\gamma & 0\\
	0 & 0 & 0 & \sin(\gamma-u)
\end{pmatrix}
\eeq
and considering $\check{R}(u)$ to act in the North-East direction, we can see that (\ref{sixvertrmat}) recovers the Boltzmann weights of the vertices in \eqref{sixvertices}. If we associate the spectral parameters $u_1$ and $u_2$ to the left and right lines as one approaches a given vertex (along the NE direction), the $\check{R}$-matrix takes the parameter $u_1-u_2$. Note that we will henceforth refer to both the $R$-matrix and the $\check{R}$-matrix, the latter being the former multiplied by a permutation operator. Consider then a square lattice where the parameters $u$ and $0$ are associated to all horizontal and vertical lines, as in figure \ref{sixvertlattice}.

\begin{figure}
	\centering
	\begin{tikzpicture}
		
\begin{scope}[thick,decoration={
    markings,
    mark=at position 0.55 with {\arrow{>}}}
    ] 
\draw[black,line width = 1pt,postaction={decorate}] (0,1)--(1,1);
\end{scope}	

\begin{scope}[thick,decoration={
    markings,
    mark=at position 0.55 with {\arrow{>}}}
    ] 
\draw[black,line width = 1pt,postaction={decorate}] (0,2)--(1,2);
\end{scope}	

\begin{scope}[thick,decoration={
    markings,
    mark=at position 0.55 with {\arrow{>}}}
    ] 
\draw[black,line width = 1pt,postaction={decorate}] (0,3)--(1,3);
\end{scope}	

\begin{scope}[thick,decoration={
    markings,
    mark=at position 0.55 with {\arrow{>}}}
    ] 
\draw[black,line width = 1pt,postaction={decorate}] (0,4)--(1,4);
\end{scope}	

\begin{scope}[thick,decoration={
    markings,
    mark=at position 0.55 with {\arrow{>}}}
    ] 
\draw[black,line width = 1pt,postaction={decorate}] (1,0)--(1,1);
\end{scope}	

\begin{scope}[thick,decoration={
    markings,
    mark=at position 0.55 with {\arrow{>}}}
    ] 
\draw[black,line width = 1pt,postaction={decorate}] (2,0)--(2,1);
\end{scope}	

\begin{scope}[thick,decoration={
    markings,
    mark=at position 0.55 with {\arrow{>}}}
    ] 
\draw[black,line width = 1pt,postaction={decorate}] (3,0)--(3,1);
\end{scope}	

\begin{scope}[thick,decoration={
    markings,
    mark=at position 0.55 with {\arrow{>}}}
    ] 
\draw[black,line width = 1pt,postaction={decorate}] (4,0)--(4,1);
\end{scope}	
		
\draw[black,line width = 1pt](1,1)--(5,1);
\draw[black,line width = 1pt](1,2)--(5,2);
\draw[black,line width = 1pt](1,3)--(5,3);
\draw[black,line width = 1pt](1,4)--(5,4);

\draw[black,line width = 1pt](1,1)--(1,5);
\draw[black,line width = 1pt](2,1)--(2,5);
\draw[black,line width = 1pt](3,1)--(3,5);
\draw[black,line width = 1pt](4,1)--(4,5);

\node at (-0.5,1) {$u$};
\node at (-0.5,2) {$u+\frac{\pi}{2}$};
\node at (-0.5,3) {$u$};
\node at (-0.5,4) {$u+\frac{\pi}{2}$};

\node at (1,-0.5) {$0$};
\node at (2,-0.5) {$\frac{\pi}{2}$};
\node at (3,-0.5) {$0$};
\node at (4,-0.5) {$\frac{\pi}{2}$};

\end{tikzpicture}
\caption{The spectral parameters on the lattice of the staggered six-vertex model .}\label{stagsixvertlattice}
\end{figure}
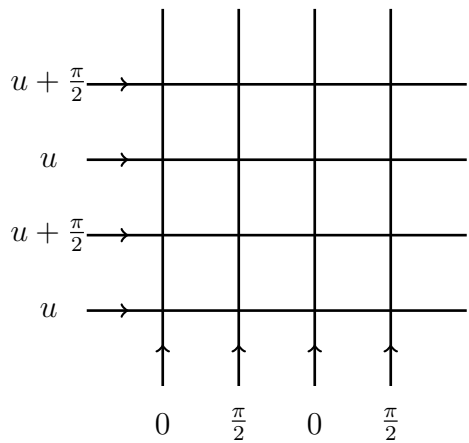

The action of $\check{R}$ on the lattice in figure \ref{sixvertlattice} recovers the correct Boltzmann weights of the six-vertex model for all of the vertices on the lattice. With this formulation of the six-vertex model we can now generalise to the staggered six-vertex model. Instead of associating $u$ and $0$ to all horizontal and vertical lines, respectively, as in figure \ref{sixvertlattice}, we will introduce a `staggering' of these parameters in both the horizontal and the vertical direction as in figure \ref{stagsixvertlattice}.

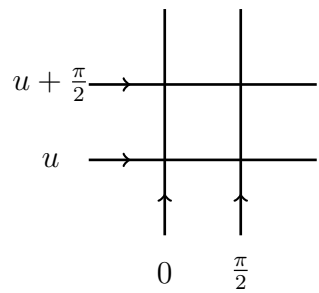
\begin{figure}
	\centering
	\begin{tikzpicture}
		
\begin{scope}[thick,decoration={
    markings,
    mark=at position 0.55 with {\arrow{>}}}
    ] 
\draw[black,line width = 1pt,postaction={decorate}] (0,1)--(1,1);
\end{scope}	

\begin{scope}[thick,decoration={
    markings,
    mark=at position 0.55 with {\arrow{>}}}
    ] 
\draw[black,line width = 1pt,postaction={decorate}] (0,2)--(1,2);
\end{scope}

\begin{scope}[thick,decoration={
    markings,
    mark=at position 0.55 with {\arrow{>}}}
    ] 
\draw[black,line width = 1pt,postaction={decorate}] (1,0)--(1,1);
\end{scope}	

\begin{scope}[thick,decoration={
    markings,
    mark=at position 0.55 with {\arrow{>}}}
    ] 
\draw[black,line width = 1pt,postaction={decorate}] (2,0)--(2,1);
\end{scope}

\draw[black,line width = 1pt](1,1)--(3,1);
\draw[black,line width = 1pt](1,2)--(3,2);

\draw[black,line width = 1pt](1,1)--(1,3);
\draw[black,line width = 1pt](2,1)--(2,3);

\node at (-0.5,1) {$u$};
\node at (-0.5,2) {$u+\frac{\pi}{2}$};

\node at (1,-0.5) {$0$};
\node at (2,-0.5) {$\frac{\pi}{2}$};

\end{tikzpicture}
\caption{The block $R$-matrix.}\label{blockrmat}
\end{figure}

This model with periodic boundary conditions was studied in detail in \cite{ikhlef2008}. The staggering can be conveniently taken into account by introducing a block $R$-matrix as in figure \ref{blockrmat}.

This new $R$-matrix now acts on the larger space

\beq
\{| \! \uparrow\uparrow\rangle, | \! \uparrow\downarrow\rangle, | \! \downarrow\uparrow\rangle, | \! \downarrow\downarrow\rangle \} \otimes \{| \! \uparrow\uparrow\rangle, | \! \uparrow\downarrow\rangle, | \! \downarrow\uparrow\rangle, | \! \downarrow\downarrow\rangle \} \,.
\eeq
As discussed in \cite{ikhlef2008}, it turns out that a convenient basis to consider is:
\beq\label{cbasis}
\{| \! \uparrow\uparrow\rangle, |0\rangle, |\bar{0}\rangle, | \! \downarrow\downarrow\rangle \} \otimes \{| \! \uparrow\uparrow\rangle, |0\rangle, |\bar{0}\rangle, | \! \downarrow\downarrow\rangle \} \,,
\eeq
where
\begin{subequations}
\begin{eqnarray}
|0 \rangle &=&\frac{1}{\sqrt{2\cos\gamma}}(e^{\frac{i\gamma}{2}}| \! \uparrow\downarrow\rangle-e^{-\frac{i\gamma}{2}}| \! \downarrow\uparrow\rangle) \,. \\
|\bar{0}\rangle &=& \frac{1}{\sqrt{2\cos\gamma}}(e^{-\frac{i\gamma}{2}}| \! \uparrow\downarrow\rangle+e^{\frac{i\gamma}{2}}| \! \downarrow\uparrow\rangle) \,.
\end{eqnarray}
\end{subequations}
In this basis there are only 38 vertices with non-zero Boltzmann weights. At each vertex, we will represent the $| \!\! \uparrow\uparrow\rangle$ state by an up- or right-pointing arrow, the $| \!\! \downarrow\downarrow\rangle$ state by a down- or left-pointing arrow, the $|0\rangle$ state by a thin line and the $|\bar{0}\rangle$ state by a thick line (the lines associated with $|0\rangle$ and $|\bar{0}\rangle$ carry no arrows). The 38 possible vertices are drawn in figure \ref{38vertices}. It was discussed in \cite{ikhlef2008} that the $R$-matrix of this 38-vertex model satisfies the Yang-Baxter equation and the model was solved via Bethe Ansatz.

Sections \ref{secstrat} and \ref{secboltz} will show that the staggered six-vertex model, or equivalently the 38-vertex model, is equivalent to the integrable model constructed from the twisted affine \dtt Lie algebra. What we mean by `equivalent' is the following: there is a well-defined procedure to start with a Lie algebra and find an $R$-matrix that satisfies the Yang-Baxter equation, and this procedure has been carried out for \dtt \cite{Jimbo1986}. When we write this $R$-matrix in a particular basis we find that there are only 38 non-zero matrix components; the \dtt $R$-matrix therefore describes a 38-vertex model. It turns out then that these matrix components are exactly those of the 38-vertex model arising from the staggered six-vertex model. \footnote{There is one subtlety that will be discussed in more detail later. Some of the matrix components of the two $R$-matrices differ by a sign, but this turns out to be unimportant because the full transfer matrix constructed from either $R$-matrix is the same.}

\subsection{Mapping between the two models: General strategy}\label{secstrat}

Starting from a given Lie algebra, one can construct an $R$-matrix that satisfies the Yang-Baxter equation. This has been carried out for the twisted affine \dtt Lie algebra in \cite{Jimbo1986}. We will show here that when written in an appropriate basis, the \dtt $R$-matrix can be identified with that of the 38-vertex model arising from the staggered six-vertex model.

The \dtt $R$-matrix is a $16\times 16$ matrix acting on the states:
\beq
\{|1\rangle,|2\rangle,|3\rangle,|4\rangle \} \otimes \{|1\rangle,|2\rangle,|3\rangle,|4\rangle \}
\eeq
where $1,2,3,4$ are just labels for the four possible states that each edge in the vertex model can take. Now define
\begin{subequations}
\begin{eqnarray}
|\tilde{2}\rangle &=& \frac{1}{\sqrt{2}}(|2\rangle+|3\rangle) \,, \\
|\tilde{3}\rangle &=& \frac{1}{\sqrt{2}}(|2\rangle-|3\rangle) \,.
\end{eqnarray}
\end{subequations}
We are interested in calculating the \dtt $R$-matrix in the basis
\beq
\{|1\rangle,|\tilde{2}\rangle,|\tilde{3}\rangle,|4\rangle \} \otimes \{|1\rangle,|\tilde{2}\rangle,|\tilde{3}\rangle,|4\rangle \} \,.
\eeq
We will do this by calculating each matrix component in the new basis term by term. The strategy is the following: first note that the \dtt $R$-matrix is written in terms of the matrices $E_{\alpha\beta} \otimes E_{\gamma\delta}$ where $E_{\alpha\beta}$ is a matrix with all components equal to zero except for the component in the $\alpha$-th row and $\beta$-th column which is equal to $1$, i.e.,
\beq
(E_{\alpha\beta})_{ij} =\delta_{i\alpha}\delta_{j\beta} \,,
\eeq
with $\alpha$,$\beta$,$\gamma$,$\delta$ taking labels $1,2,3,$ or $4$. To calculate the matrix elements in the new basis we need to expand the $R$-matrix in terms of matrices $E_{\tilde{\alpha}\tilde{\beta}} \otimes E_{\tilde{\gamma}\tilde{\delta}}$ where $\tilde{\alpha},\tilde{\beta},\tilde{\gamma},\tilde{\delta}$ take labels 1,$\tilde{2},\tilde{3}$ or $4$ and we have
\begin{subequations}
\begin{eqnarray}
E_{\tilde{\alpha}\tilde{2}} &=& \frac{1}{\sqrt{2}}(E_{\tilde{\alpha}2}+E_{\tilde{\alpha}3}) \,, \\
E_{\tilde{2}\tilde{\alpha}} &=& \frac{1}{\sqrt{2}}(E_{2\tilde{\alpha}}+E_{3\tilde{\alpha}}) \,, \\
E_{\tilde{\alpha}\tilde{3}} &=& \frac{1}{\sqrt{2}}(E_{\tilde{\alpha}2}-E_{\tilde{\alpha}3}) \,, \\
E_{\tilde{3}\tilde{\alpha}} &=& \frac{1}{\sqrt{2}}(E_{2\tilde{\alpha}}-E_{3\tilde{\alpha}}) \,.
\end{eqnarray}
\end{subequations}
We have then, for example:
\begin{subequations}
\label{newbasisex}
\begin{eqnarray}
E_{1\tilde{2}}\otimes E_{1\tilde{2}} &=& \frac{1}{2}(E_{12}\otimes E_{12}+E_{12}\otimes E_{13}+E_{13}\otimes E_{12}+E_{13}\otimes E_{13})  \,, \\
E_{1\tilde{2}}\otimes E_{\tilde{2}\tilde{3}} &=& \frac{1}{2\sqrt{2}} (E_{12}\otimes E_{22}+E_{12}\otimes E_{32}-E_{12}\otimes E_{23}-E_{12}\otimes E_{33} \nonumber \\
& & \qquad + E_{13}\otimes E_{22}+E_{13}\otimes E_{32}-E_{13}\otimes E_{23}-E_{13}\otimes E_{33}) \,.
\end{eqnarray}
\end{subequations}
When we expand the \dtt $R$-matrix in terms of the matrices $E_{\tilde{\alpha}\tilde{\beta}} \otimes E_{\tilde{\gamma}\tilde{\delta}}$, the coefficient of each of these terms will give the Boltzmann weight of exactly one vertex. It will turn out that, in this basis, there are exactly 38 non-zero coefficients and that these coefficients are the Boltzmann weights of the 38-vertex model arising from staggered six-vertex model. The coefficient in front of the term in (\ref{newbasisex}) will correspond to the Boltzmann weight of one of these 38 vertices, as we now discuss.

\subsection{Deriving the Boltzmann weights}\label{secboltz}

\begin{figure}
	\centering
	\begin{tikzpicture}
\draw[black,line width = 1pt](0,1)--(0.5,1);
\draw[black,line width = 1pt](0.5,1)--(0.5,0.5);
\draw[black,line width = 1pt](0.5,1)--(1,1);
\draw[black,line width = 1pt](0.5,1)--(0.5,1.5);

\node at (0.43,0.3) {$\delta$};
\node at (0.44,1.7) {$\gamma$};
\node at (-0.2,1) {$\beta$};
\node at (1.17,1) {$\alpha$};

\end{tikzpicture}
\caption{Labelling around a vertex in the \dtt model.}\label{genericvertex}
\end{figure}
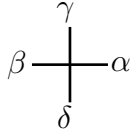

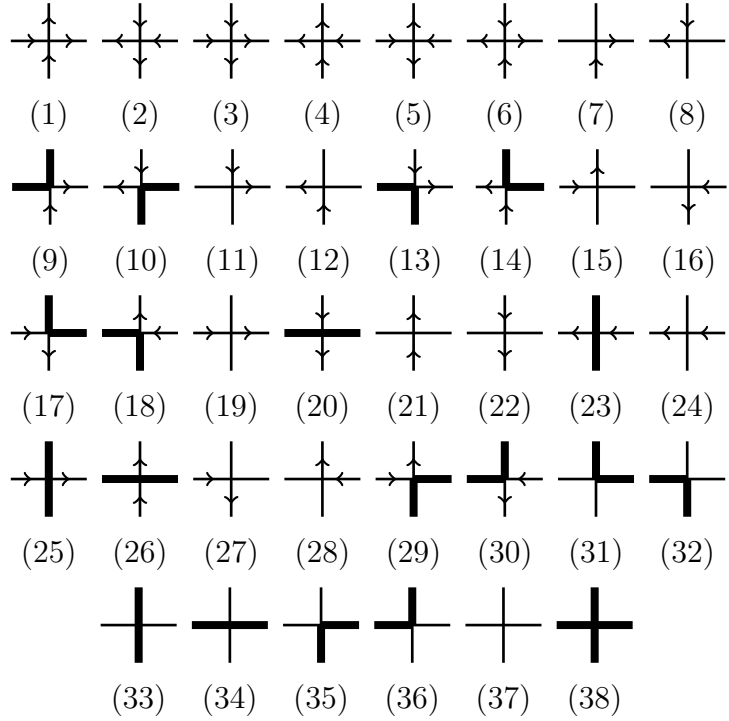
\begin{figure}
	\centering
	\begin{tikzpicture}
\begin{scope}[thick,decoration={
    markings,
    mark=at position 0.65 with {\arrow{>}}}
    ] 
\draw[black,line width = 1pt,postaction={decorate}] (0,1)--(0.5,1);
\draw[black,line width = 1pt,postaction={decorate}](0.5,0.5)--(0.5,1);
\draw[black,line width = 1pt,postaction={decorate}](0.5,1)--(1,1);
\draw[black,line width = 1pt,postaction={decorate}](0.5,1)--(0.5,1.5);
\end{scope}

\node at (0.5,0) {(1)};

\begin{scope}[thick,decoration={
    markings,
    mark=at position 0.55 with {\arrow{<}}}
    ] 
\draw[black,line width = 1pt,postaction={decorate}] (1.7,1)--(2.2,1);
\draw[black,line width = 1pt,postaction={decorate}] (1.7,0.5)--(1.7,1);
\draw[black,line width = 1pt,postaction={decorate}] (1.2,1)--(1.7,1);
 \draw[black,line width = 1pt,postaction={decorate}](1.7,1)--(1.7,1.5);
\end{scope}

\node at (1.7,0) {(2)};

\begin{scope}[thick,decoration={
    markings,
    mark=at position 0.55 with {\arrow{<}}}
    ] 

\draw[black,line width = 1pt,postaction={decorate}] (2.9,0.5)--(2.9,1);
 \draw[black,line width = 1pt,postaction={decorate}](2.9,1)--(2.9,1.5);
\end{scope}

\begin{scope}[thick,decoration={
    markings,
    mark=at position 0.55 with {\arrow{>}}}
    ] 
\draw[black,line width = 1pt,postaction={decorate}] (2.9,1)--(3.4,1);
\draw[black,line width = 1pt,postaction={decorate}] (2.4,1)--(2.9,1);
\end{scope}

\node at (2.9,0) {(3)};

\begin{scope}[thick,decoration={
    markings,
    mark=at position 0.55 with {\arrow{<}}}
    ] 

\draw[black,line width = 1pt,postaction={decorate}] (3.6,1)--(4.1,1);
 \draw[black,line width = 1pt,postaction={decorate}](4.1,1)--(4.6,1);
\end{scope}

\begin{scope}[thick,decoration={
    markings,
    mark=at position 0.55 with {\arrow{>}}}
    ] 
\draw[black,line width = 1pt,postaction={decorate}] (4.1,0.5)--(4.1,1);
\draw[black,line width = 1pt,postaction={decorate}] (4.1,1)--(4.1,1.5);
\end{scope}

\node at (4.1,0) {(4)};

\begin{scope}[thick,decoration={
    markings,
    mark=at position 0.55 with {\arrow{<}}}
    ] 

\draw[black,line width = 1pt,postaction={decorate}] (5.3,1)--(5.8,1);
 \draw[black,line width = 1pt,postaction={decorate}](5.3,0.5)--(5.3,1);
\end{scope}

\begin{scope}[thick,decoration={
    markings,
    mark=at position 0.55 with {\arrow{>}}}
    ] 
\draw[black,line width = 1pt,postaction={decorate}] (4.8,1)--(5.3,1);
\draw[black,line width = 1pt,postaction={decorate}] (5.3,1)--(5.3,1.5);
\end{scope}

\node at (5.3,0) {(5)};

\begin{scope}[thick,decoration={
    markings,
    mark=at position 0.55 with {\arrow{<}}}
    ] 

\draw[black,line width = 1pt,postaction={decorate}] (6,1)--(6.5,1);
 \draw[black,line width = 1pt,postaction={decorate}](6.5,1)--(6.5,1.5);
\end{scope}

\begin{scope}[thick,decoration={
    markings,
    mark=at position 0.55 with {\arrow{>}}}
    ] 
\draw[black,line width = 1pt,postaction={decorate}] (6.5,0.5)--(6.5,1);
\draw[black,line width = 1pt,postaction={decorate}] (6.5,1)--(7,1);
\end{scope}

\node at (6.5,0) {(6)};

\begin{scope}[thick,decoration={
    markings,
    mark=at position 0.55 with {\arrow{>}}}
    ] 

\draw[black,line width = 1pt,postaction={decorate}] (7.7,0.5)--(7.7,1);
 \draw[black,line width = 1pt,postaction={decorate}](7.7,1)--(8.2,1);
\end{scope}

\draw[black,line width = 1pt,postaction={decorate}] (7.2,1)--(7.7,1);
\draw[black,line width = 1pt,postaction={decorate}] (7.7,1)--(7.7,1.5);

\node at (7.7,0) {(7)};

\begin{scope}[thick,decoration={
    markings,
    mark=at position 0.55 with {\arrow{<}}}
    ] 

\draw[black,line width = 1pt,postaction={decorate}] (8.4,1)--(8.9,1);
 \draw[black,line width = 1pt,postaction={decorate}](8.9,1)--(8.9,1.5);
\end{scope}

\draw[black,line width = 1pt,postaction={decorate}] (8.9,0.5)--(8.9,1);
\draw[black,line width = 1pt,postaction={decorate}] (8.9,1)--(9.4,1);

\node at (8.9,0) {(8)};
\end{tikzpicture}\\
\begin{tikzpicture}

\begin{scope}[thick,decoration={
    markings,
    mark=at position 0.55 with {\arrow{>}}}
    ] 

\draw[black,line width = 1pt,postaction={decorate}] (0.5,0.5)--(0.5,1);
 \draw[black,line width = 1pt,postaction={decorate}](0.5,1)--(1,1);
\end{scope}

\draw[black,line width = 3pt,postaction={decorate}] (0,1)--(0.5,1);
\draw[black,line width = 3pt,postaction={decorate}] (0.5,1)--(0.5,1.5);

\node at (0.5,0) {(9)};

\begin{scope}[thick,decoration={
    markings,
    mark=at position 0.55 with {\arrow{<}}}
    ] 

\draw[black,line width = 1pt,postaction={decorate}] (1.2,1)--(1.7,1);
 \draw[black,line width = 1pt,postaction={decorate}](1.7,1)--(1.7,1.5);
\end{scope}

\draw[black,line width = 3pt,postaction={decorate}] (1.7,0.5)--(1.7,1);
\draw[black,line width = 3pt,postaction={decorate}] (1.7,1)--(2.2,1);

\node at (1.7,0) {(10)};

\begin{scope}[thick,decoration={
    markings,
    mark=at position 0.55 with {\arrow{>}}}
    ] 
\draw[black,line width = 1pt,postaction={decorate}] (2.9,1)--(3.4,1);
\end{scope}

\begin{scope}[thick,decoration={
    markings,
    mark=at position 0.55 with {\arrow{<}}}
    ] 
\draw[black,line width = 1pt,postaction={decorate}] (2.9,1)--(2.9,1.5);
\end{scope}

\draw[black,line width = 1pt,postaction={decorate}] (2.4,1)--(2.9,1);
\draw[black,line width = 1pt,postaction={decorate}] (2.9,0.5)--(2.9,1);

\node at (2.9,0) {(11)};

\begin{scope}[thick,decoration={
    markings,
    mark=at position 0.55 with {\arrow{>}}}
    ] 
\draw[black,line width = 1pt,postaction={decorate}] (4.1,0.5)--(4.1,1);
\end{scope}

\begin{scope}[thick,decoration={
    markings,
    mark=at position 0.55 with {\arrow{<}}}
    ] 
\draw[black,line width = 1pt,postaction={decorate}] (3.6,1)--(4.1,1);
\end{scope}

\draw[black,line width = 1pt,postaction={decorate}] (4.1,1)--(4.6,1);
\draw[black,line width = 1pt,postaction={decorate}] (4.1,1)--(4.1,1.5);

\node at (4.1,0) {(12)};

\begin{scope}[thick,decoration={
    markings,
    mark=at position 0.55 with {\arrow{>}}}
    ] 
\draw[black,line width = 1pt,postaction={decorate}] (5.3,1)--(5.8,1);
\end{scope}

\begin{scope}[thick,decoration={
    markings,
    mark=at position 0.55 with {\arrow{<}}}
    ] 
\draw[black,line width = 1pt,postaction={decorate}] (5.3,1)--(5.3,1.5);
\end{scope}

\draw[black,line width = 3pt,postaction={decorate}] (4.8,1)--(5.3,1);
\draw[black,line width = 3pt,postaction={decorate}] (5.3,0.5)--(5.3,1);

\node at (5.3,0) {(13)};

\begin{scope}[thick,decoration={
    markings,
    mark=at position 0.55 with {\arrow{>}}}
    ] 
\draw[black,line width = 1pt,postaction={decorate}] (6.5,0.5)--(6.5,1);
\end{scope}

\begin{scope}[thick,decoration={
    markings,
    mark=at position 0.55 with {\arrow{<}}}
    ] 
\draw[black,line width = 1pt,postaction={decorate}] (6.1,1)--(6.5,1);
\end{scope}

\draw[black,line width = 3pt,postaction={decorate}] (6.5,1)--(7,1);
\draw[black,line width = 3pt,postaction={decorate}] (6.5,1)--(6.5,1.5);

\node at (6.5,0) {(14)};

\begin{scope}[thick,decoration={
    markings,
    mark=at position 0.55 with {\arrow{>}}}
    ] 
\draw[black,line width = 1pt,postaction={decorate}] (7.2,1)--(7.7,1);
\end{scope}

\begin{scope}[thick,decoration={
    markings,
    mark=at position 0.55 with {\arrow{>}}}
    ] 
\draw[black,line width = 1pt,postaction={decorate}] (7.7,1)--(7.7,1.5);
\end{scope}

\draw[black,line width = 1pt,postaction={decorate}] (7.7,0.5)--(7.7,1);
\draw[black,line width = 1pt,postaction={decorate}] (7.7,1)--(8.2,1);

\node at (7.7,0) {(15)};

\begin{scope}[thick,decoration={
    markings,
    mark=at position 0.55 with {\arrow{<}}}
    ] 
\draw[black,line width = 1pt,postaction={decorate}] (8.9,0.5)--(8.9,1);
\end{scope}

\begin{scope}[thick,decoration={
    markings,
    mark=at position 0.55 with {\arrow{<}}}
    ] 
\draw[black,line width = 1pt,postaction={decorate}] (8.9,1)--(9.4,1);
\end{scope}

\draw[black,line width = 1pt,postaction={decorate}] (8.4,1)--(8.9,1);
\draw[black,line width = 1pt,postaction={decorate}] (8.9,1)--(8.9,1.5);

\node at (8.9,0) {(16)};
\end{tikzpicture}\\

\begin{tikzpicture}

\begin{scope}[thick,decoration={
    markings,
    mark=at position 0.55 with {\arrow{>}}}
    ] 
\draw[black,line width = 1pt,postaction={decorate}] (0,1)--(0.5,1);
\end{scope}

\begin{scope}[thick,decoration={
    markings,
    mark=at position 0.55 with {\arrow{<}}}
    ] 
\draw[black,line width = 1pt,postaction={decorate}] (0.5,0.5)--(0.5,1);
\end{scope}

\draw[black,line width = 3pt,postaction={decorate}] (0.5,1)--(1,1);
\draw[black,line width = 3pt,postaction={decorate}] (0.5,1)--(0.5,1.5);

\node at (0.5,0) {(17)};

\begin{scope}[thick,decoration={
    markings,
    mark=at position 0.55 with {\arrow{<}}}
    ] 
\draw[black,line width = 1pt,postaction={decorate}] (1.7,1)--(2.2,1);
\end{scope}

\begin{scope}[thick,decoration={
    markings,
    mark=at position 0.55 with {\arrow{>}}}
    ] 
\draw[black,line width = 1pt,postaction={decorate}] (1.7,1)--(1.7,1.5);
\end{scope}

\draw[black,line width = 3pt,postaction={decorate}] (1.2,1)--(1.7,1);
\draw[black,line width = 3pt,postaction={decorate}] (1.7,0.5)--(1.7,1);

\node at (1.7,0) {(18)};

\begin{scope}[thick,decoration={
    markings,
    mark=at position 0.55 with {\arrow{>}}}
    ] 
\draw[black,line width = 1pt,postaction={decorate}] (2.4,1)--(2.9,1);
\end{scope}

\begin{scope}[thick,decoration={
    markings,
    mark=at position 0.55 with {\arrow{>}}}
    ] 
\draw[black,line width = 1pt,postaction={decorate}] (2.9,1)--(3.4,1);
\end{scope}

\draw[black,line width = 1pt,postaction={decorate}] (2.9,0.5)--(2.9,1);
\draw[black,line width = 1pt,postaction={decorate}] (2.9,1)--(2.9,1.5);

\node at (2.9,0) {(19)};

\begin{scope}[thick,decoration={
    markings,
    mark=at position 0.55 with {\arrow{<}}}
    ] 
\draw[black,line width = 1pt,postaction={decorate}] (4.1,0.5)--(4.1,1);
\end{scope}

\begin{scope}[thick,decoration={
    markings,
    mark=at position 0.55 with {\arrow{<}}}
    ] 
\draw[black,line width = 1pt,postaction={decorate}] (4.1,1)--(4.1,1.5);
\end{scope}

\draw[black,line width = 3pt,postaction={decorate}] (3.6,1)--(4.1,1);
\draw[black,line width = 3pt,postaction={decorate}] (4.1,1)--(4.6,1);

\node at (4.1,0) {(20)};

\begin{scope}[thick,decoration={
    markings,
    mark=at position 0.55 with {\arrow{>}}}
    ] 
\draw[black,line width = 1pt,postaction={decorate}] (5.3,0.5)--(5.3,1);
\end{scope}

\begin{scope}[thick,decoration={
    markings,
    mark=at position 0.55 with {\arrow{>}}}
    ] 
\draw[black,line width = 1pt,postaction={decorate}] (5.3,1)--(5.3,1.5);
\end{scope}

\draw[black,line width = 1pt,postaction={decorate}] (4.8,1)--(5.3,1);
\draw[black,line width = 1pt,postaction={decorate}] (5.3,1)--(5.8,1);

\node at (5.3,0) {(21)};

\begin{scope}[thick,decoration={
    markings,
    mark=at position 0.55 with {\arrow{<}}}
    ] 
\draw[black,line width = 1pt,postaction={decorate}] (6.5,0.5)--(6.5,1);
\end{scope}

\begin{scope}[thick,decoration={
    markings,
    mark=at position 0.55 with {\arrow{<}}}
    ] 
\draw[black,line width = 1pt,postaction={decorate}] (6.5,1)--(6.5,1.5);
\end{scope}

\draw[black,line width = 1pt,postaction={decorate}] (6,1)--(6.5,1);
\draw[black,line width = 1pt,postaction={decorate}] (6.5,1)--(7,1);

\node at (6.5,0) {(22)};

\begin{scope}[thick,decoration={
    markings,
    mark=at position 0.55 with {\arrow{<}}}
    ] 
\draw[black,line width = 1pt,postaction={decorate}] (7.2,1)--(7.7,1);
\end{scope}

\begin{scope}[thick,decoration={
    markings,
    mark=at position 0.55 with {\arrow{<}}}
    ] 
\draw[black,line width = 1pt,postaction={decorate}] (7.7,1)--(8.2,1);
\end{scope}

\draw[black,line width = 3pt,postaction={decorate}] (7.7,0.5)--(7.7,1);
\draw[black,line width = 3pt,postaction={decorate}] (7.7,1)--(7.7,1.5);

\node at (7.7,0) {(23)};

\begin{scope}[thick,decoration={
    markings,
    mark=at position 0.55 with {\arrow{<}}}
    ] 
\draw[black,line width = 1pt,postaction={decorate}] (8.4,1)--(8.9,1);
\end{scope}

\begin{scope}[thick,decoration={
    markings,
    mark=at position 0.55 with {\arrow{<}}}
    ] 
\draw[black,line width = 1pt,postaction={decorate}] (8.9,1)--(9.4,1);
\end{scope}

\draw[black,line width = 1pt,postaction={decorate}] (8.9,0.5)--(8.9,1);
\draw[black,line width = 1pt,postaction={decorate}] (8.9,1)--(8.9,1.5);

\node at (8.9,0) {(24)};
\end{tikzpicture}\\

\begin{tikzpicture}
	\begin{scope}[thick,decoration={
	    markings,
	    mark=at position 0.55 with {\arrow{>}}}
	    ] 
	\draw[black,line width = 1pt,postaction={decorate}] (0,1)--(0.5,1);
	\end{scope}

	\begin{scope}[thick,decoration={
	    markings,
	    mark=at position 0.55 with {\arrow{>}}}
	    ] 
	\draw[black,line width = 1pt,postaction={decorate}] (0.5,1)--(1,1);
	\end{scope}

	\draw[black,line width = 3pt,postaction={decorate}] (0.5,0.5)--(0.5,1);
	\draw[black,line width = 3pt,postaction={decorate}] (0.5,1)--(0.5,1.5);

	\node at (0.5,0) {(25)};
	
	\begin{scope}[thick,decoration={
	    markings,
	    mark=at position 0.55 with {\arrow{>}}}
	    ] 
	\draw[black,line width = 1pt,postaction={decorate}] (1.7,0.5)--(1.7,1);
	\end{scope}

	\begin{scope}[thick,decoration={
	    markings,
	    mark=at position 0.55 with {\arrow{>}}}
	    ] 
	\draw[black,line width = 1pt,postaction={decorate}] (1.7,1)--(1.7,1.5);
	\end{scope}

	\draw[black,line width = 3pt,postaction={decorate}] (1.2,1)--(1.7,1);
	\draw[black,line width = 3pt,postaction={decorate}] (1.7,1)--(2.2,1);

	\node at (1.7,0) {(26)};
	
	\begin{scope}[thick,decoration={
	    markings,
	    mark=at position 0.55 with {\arrow{>}}}
	    ] 
	\draw[black,line width = 1pt,postaction={decorate}] (2.4,1)--(2.9,1);
	\end{scope}

	\begin{scope}[thick,decoration={
	    markings,
	    mark=at position 0.55 with {\arrow{<}}}
	    ] 
	\draw[black,line width = 1pt,postaction={decorate}] (2.9,0.5)--(2.9,1);
	\end{scope}

	\draw[black,line width = 1pt,postaction={decorate}] (2.9,1)--(3.4,1);
	\draw[black,line width = 1pt,postaction={decorate}] (2.9,1)--(2.9,1.5);

	\node at (2.9,0) {(27)};
	
	\begin{scope}[thick,decoration={
	    markings,
	    mark=at position 0.55 with {\arrow{>}}}
	    ] 
	\draw[black,line width = 1pt,postaction={decorate}] (4.1,1)--(4.1,1.5);
	\end{scope}

	\begin{scope}[thick,decoration={
	    markings,
	    mark=at position 0.55 with {\arrow{<}}}
	    ] 
	\draw[black,line width = 1pt,postaction={decorate}] (4.1,1)--(4.6,1);
	\end{scope}

	\draw[black,line width = 1pt,postaction={decorate}] (3.6,1)--(4.1,1);
	\draw[black,line width = 1pt,postaction={decorate}] (4.1,0.5)--(4.1,1);

	\node at (4.1,0) {(28)};
	
	\begin{scope}[thick,decoration={
	    markings,
	    mark=at position 0.55 with {\arrow{>}}}
	    ] 
	\draw[black,line width = 1pt,postaction={decorate}] (4.8,1)--(5.3,1);
	\end{scope}

	\begin{scope}[thick,decoration={
	    markings,
	    mark=at position 0.55 with {\arrow{>}}}
	    ] 
	\draw[black,line width = 1pt,postaction={decorate}] (5.3,1)--(5.3,1.5);
	\end{scope}

	\draw[black,line width = 3pt,postaction={decorate}] (5.3,0.5)--(5.3,1);
	\draw[black,line width = 3pt,postaction={decorate}] (5.3,1)--(5.8,1);

	\node at (5.3,0) {(29)};

	\begin{scope}[thick,decoration={
	    markings,
	    mark=at position 0.55 with {\arrow{<}}}
	    ] 
	\draw[black,line width = 1pt,postaction={decorate}] (6.5,0.5)--(6.5,1);
	\end{scope}

	\begin{scope}[thick,decoration={
	    markings,
	    mark=at position 0.55 with {\arrow{<}}}
	    ] 
	\draw[black,line width = 1pt,postaction={decorate}] (6.5,1)--(7,1);
	\end{scope}

	\draw[black,line width = 3pt,postaction={decorate}] (6,1)--(6.5,1);
	\draw[black,line width = 3pt,postaction={decorate}] (6.5,1)--(6.5,1.5);

	\node at (6.5,0) {(30)};	

	\draw[black,line width = 1pt,postaction={decorate}] (7.2,1)--(7.7,1);
	\draw[black,line width = 1pt,postaction={decorate}] (7.7,0.5)--(7.7,1);
	\draw[black,line width = 3pt,postaction={decorate}] (7.7,1)--(8.2,1);
	\draw[black,line width = 3pt,postaction={decorate}] (7.7,1)--(7.7,1.5);

	\node at (7.7,0) {(31)};

	\draw[black,line width = 1pt,postaction={decorate}] (8.9,1)--(9.4,1);
	\draw[black,line width = 1pt,postaction={decorate}] (8.9,1)--(8.9,1.5);
	\draw[black,line width = 3pt,postaction={decorate}] (8.4,1)--(8.9,1);
	\draw[black,line width = 3pt,postaction={decorate}] (8.9,0.5)--(8.9,1);

	\node at (8.9,0) {(32)};
	
\end{tikzpicture}\\

\begin{tikzpicture}
	\draw[black,line width = 1pt,postaction={decorate}] (0,1)--(0.5,1);
	\draw[black,line width = 1pt,postaction={decorate}] (0.5,1)--(1,1);
	\draw[black,line width = 3pt,postaction={decorate}] (0.5,0.5)--(0.5,1);
	\draw[black,line width = 3pt,postaction={decorate}] (0.5,1)--(0.5,1.5);

	\node at (0.5,0) {(33)};
	
	\draw[black,line width = 1pt,postaction={decorate}] (1.7,0.5)--(1.7,1);
	\draw[black,line width = 1pt,postaction={decorate}] (1.7,1)--(1.7,1.5);
	\draw[black,line width = 3pt,postaction={decorate}] (1.2,1)--(1.7,1);
	\draw[black,line width = 3pt,postaction={decorate}] (1.7,1)--(2.2,1);

	\node at (1.7,0) {(34)};
	
	\draw[black,line width = 1pt,postaction={decorate}] (2.4,1)--(2.9,1);
	\draw[black,line width = 1pt,postaction={decorate}] (2.9,1)--(2.9,1.5);
	\draw[black,line width = 3pt,postaction={decorate}] (2.9,0.5)--(2.9,1);
	\draw[black,line width = 3pt,postaction={decorate}] (2.9,1)--(3.4,1);

	\node at (2.9,0) {(35)};
	
	\draw[black,line width = 1pt,postaction={decorate}] (4.1,0.5)--(4.1,1);
	\draw[black,line width = 1pt,postaction={decorate}] (4.1,1)--(4.6,1);
	\draw[black,line width = 3pt,postaction={decorate}] (3.6,1)--(4.1,1);
	\draw[black,line width = 3pt,postaction={decorate}] (4.1,1)--(4.1,1.5);

	\node at (4.1,0) {(36)};
	
	\draw[black,line width = 1pt,postaction={decorate}] (4.8,1)--(5.3,1);
	\draw[black,line width = 1pt,postaction={decorate}] (5.3,0.5)--(5.3,1);
	\draw[black,line width = 1pt,postaction={decorate}] (5.3,1)--(5.8,1);
	\draw[black,line width = 1pt,postaction={decorate}] (5.3,1)--(5.3,1.5);

	\node at (5.3,0) {(37)};
	
	\draw[black,line width = 3pt,postaction={decorate}] (6,1)--(6.5,1);
	\draw[black,line width = 3pt,postaction={decorate}] (6.5,0.5)--(6.5,1);
	\draw[black,line width = 3pt,postaction={decorate}] (6.5,1)--(7,1);
	\draw[black,line width = 3pt,postaction={decorate}] (6.5,1)--(6.5,1.5);

	\node at (6.5,0) {(38)};

\end{tikzpicture}

\caption{The 38 vertices of the \dtt model.}\label{38vertices}
\end{figure}

The \dtt $R$-matrix is expanded in terms of the matrices $E_{\alpha\beta} \otimes E_{\gamma\delta}$:

\beq\label{rmatexp}
R=\sum\limits_{\alpha\beta\gamma\delta}\omega_{\alpha\beta\gamma\delta}E_{\alpha\beta}\otimes E_{\gamma\delta} \,.
\eeq
We then interpret $\omega_{\alpha\beta\gamma\delta}$ as the Boltzmann weight of the vertex in figure \ref{genericvertex}. In particular, $\alpha$ is the label of the state of the right edge, $\beta$ the label of the state on the left edge, $\gamma$ the label of the state on the top edge and $\delta$ the label of the state on the bottom edge. We will represent these labels in the following way: associate a down- or left-pointing arrow to the label $1$, an up- or right-pointing arrow the label $4$, a thin line to the label $\tilde{2}$ and a thick line to the label $\tilde{3}$. The coefficient $\omega_{1111}$ for example then gives the Boltzmann weight of vertex (1) in figure \ref{38vertices}, and the coefficient $\omega_{4\tilde{3}1\tilde{3}}$ gives the Boltzmann weight of vertex (13).

The explicit expression for the $R$-matrix of the \dtt model can be found in equation (3.7) of \cite{Jimbo1986}. The important point for us is that this expression for the \dtt $R$-matrix is of the form (\ref{rmatexp}) and that the weights $\omega_{\alpha\beta\gamma\delta}$ are written in terms of parameters $x$ and $k$. Meanwhile, the explicit expression for the 38-vertex model can be found in section 2.3.3 of \cite{ikhlef2008} and this matrix is written in terms of parameters $u_0$ and $\gamma_0$. The latter two parameters are related to those of the six-vertex model $R$-matrix \eqref{sixvertrmat} by \cite{ikhlef2008}
\begin{subequations}
\label{u0gam0}
\begin{eqnarray}
u_0 &=& -2u \,, \\
\gamma_0 &=& \pi-2\gamma \,.
\end{eqnarray}
\end{subequations}
It will turn out that the correct associations between the parameters of the two models are
\begin{subequations}
\label{kx_corr}
\begin{eqnarray}
k &=& -e^{-i\gamma_0} \,, \\
x &=& e^{-iu_0} \,,
\end{eqnarray}
\end{subequations}
so that we have 
\begin{subequations}
\begin{eqnarray}
k &=& e^{2i\gamma} \,, \\
x &=& e^{2iu} \,.
\end{eqnarray}
\end{subequations}
With this identification, our goal is then to write the $R$-matrix in a new basis:
\beq
\label{rnew}
R=\sum\limits_{\tilde{\alpha}\tilde{\beta}\tilde{\gamma}\tilde{\delta}}\tilde{\omega}_{\tilde{\alpha}\tilde{\beta}\tilde{\gamma}\tilde{\delta}}E_{\tilde{\alpha}\tilde{\beta}}\otimes E_{\tilde{\gamma}\tilde{\delta}} \,,
\eeq
and to calculate the weights $\tilde{\omega}_{\tilde{\alpha}\tilde{\beta}\tilde{\gamma}\tilde{\delta}}$ in terms of the parameters $u_0$ and $\gamma_0$ by writing $R$ in the form \eqref{rnew}. Consider all the vertices of the 38-vertex model in figure \ref{38vertices}. There are three types of vertices to consider: vertices with four arrows (1 to 6), two arrows (7 to 30) and no arrows (31 to 38). We will study each of these three types of vertices individually.

\subsubsection{Vertices 1 to 6}\label{secvert1to6}
These vertices have four arrows (two in and two out). Since we associate arrows with states $|1\rangle$ and $|4\rangle$ there will be no change to the Boltzmann weights of these vertices when we change from the old basis $|1\rangle$,$|2\rangle$,$|3\rangle$,$|4\rangle$ to the new basis $|1\rangle$,$|\tilde{2}\rangle$,$|\tilde{3}\rangle$,$|4\rangle$, except for the change in parameters from $x$ and $k$ to $u_0$ and $\gamma_0$. Consider for example vertices 1 and 2. These correspond to the terms $\omega_{4444}E_{44}\otimes E_{44}$ and $\omega_{1111}E_{11}\otimes E_{11}$ in the expansion of the $R$-matrix. We have from \cite{Jimbo1986}:
\beq
\omega_{1111}=\omega_{4444}=(x^2-k^2)^2 \,,
\eeq
and we know that $\omega_{1111}=\tilde{\omega}_{1111}$ and $\omega_{4444}=\tilde{\omega}_{4444}$.
Using \eqref{kx_corr} we then find
\beq
\omega_{1111}=\omega_{4444}=-4k^2 x^2\sin^2(\gamma_0-u_0) \,,
\eeq
which is equal to the weight of these vertices in the staggered six-vertex model, up to an overall factor of $16 k^2x^2$ which will turn out to be present in all terms. We can perform the same calculation for vertices 3 to 7, the results of which are shown in table \ref{vert1to6}. We see that when we make the associations $x=\exp(-iu_0)$ and $k=-\exp(-i\gamma_0)$, as in \eqref{kx_corr}, all of these vertices have the same Boltzmann weight in the \dtt model and the staggered six-vertex model, again up to a factor of $16 k^2 x^2$.

\begin{table}
\begin{center} 
\begin{tabular}{|c|c|c|c}
\hline
Vertex  & \dtt weight & Staggered six-vertex weight\\
\hline
1 & $-4k^2x^2\sin^2(\gamma_0 -u_0)$ & $-\frac{1}{4}\sin^2(\gamma_0 -u_0)$\\
\hline
2 & $-4k^2x^2\sin^2(\gamma_0 -u_0)$ & $-\frac{1}{4}\sin^2(\gamma_0 -u_0)$\\
\hline
3 & $-4k^2x^2\sin^2(u_0)$ & $-\frac{1}{4}\sin^2(u_0)$\\
\hline
4 & $-4k^2x^2\sin^2(u_0)$ & $-\frac{1}{4}\sin^2(u_0)$\\
\hline
5 & $4k^2x^2 e^{-2iu}\sin(\gamma_0)(\sin(u_0)-\sin(\gamma_0-u_0))$ & $\frac{1}{4}e^{-2iu}\sin(\gamma_0)(\sin(u_0)-\sin(\gamma_0-u_0))$\\
\hline
6 & $4k^2x^2 e^{2iu}\sin(\gamma_0)(\sin(u_0)-\sin(\gamma_0-u_0))$ & $\frac{1}{4}e^{2iu}\sin(\gamma_0)(\sin(u_0)-\sin(\gamma_0-u_0))$\\
\hline
\end{tabular}
\caption{Correspondence between the Boltzmann weights of the \dtt model and the staggered six-vertex model. Vertices 1 to 6.}\label{vert1to6}
\end{center}
\end{table}

\subsubsection{Vertices 7 to 30}\label{secvert7to30}
\no We will now show an example of a calculation of the \dtt Boltzmann weight of a vertex with two arrows. Consider vertices 8 and 10, which correspond to the terms $E_{\tilde{2}1} \otimes E_{1\tilde{2}}$ and $E_{\tilde{3}1}\otimes E_{1\tilde{3}}$ in the expansion of the $R$-matrix. We will calculate the coefficients of these terms when we change basis from $|1\rangle,|2\rangle,|3\rangle,|4\rangle$ to $|1\rangle,|\tilde{2}\rangle,|\tilde{3}\rangle,|4\rangle$. Consider the following terms appearing in the expansion of the $R$-matrix in the old basis:
\beq
-\frac{1}{2}(k^2-1)(x^2-k^2)(x+1)x(E_{21}\otimes E_{12}+E_{31}\otimes E_{13})-\frac{1}{2}x(k^2-1)(x^2-k^2)(x-1)(E_{21}\otimes E_{13}+E_{31}\otimes E_{12}) \,.
\eeq
This can be reformulated as
\beq
-\frac{1}{2}(k^2-1)(x^2-k^2)x[x(E_{21}+E_{31})\otimes E_{12}+(E_{21}-E_{31})\otimes E_{12}+x(E_{21}+E_{31})\otimes E_{13}-(E_{21}-E_{31})\otimes E_{13}] \,,
\eeq
which we can see gives:
\beq
-(k^2-1)(x^2-k^2)x[xE_{\tilde{2}1}\otimes E_{1\tilde{2}}+E_{\tilde{3}1}\otimes E_{1\tilde{3}}] \,.
\eeq
After making the associations \eqref{kx_corr} we finally obtain
\beq\label{vert7and9}
-4 k^2 x^2 e^{2iu}\sin(\gamma_0-u_0)\sin(\gamma_0)[E_{\tilde{2}1}\otimes E_{1\tilde{2}}]+4 k^2 x^2 \sin(\gamma_0-u_0)\sin(\gamma_0)[E_{\tilde{3}1}\otimes E_{1\tilde{3}}] \,.
\eeq
The coefficients of the two terms in (\ref{vert7and9}) give the Boltzmann weights of vertices 8 and 10 in figure \ref{38vertices} and are compared with the Boltzmann weights of the staggered six-vertex model in table \ref{vert7to30}. We observe that the Boltzmann weights in the two models are equal, again up to the factor of $16 k^2 x^2$. In the case of vertex 10, there is a difference of sign between the two models. Vertices with a sign difference in the two models are marked with an asterisk in the last column of the table. This sign difference will turn out not to affect the transfer matrix built from the $R$-matrix and therefore not to have any effect on the physics. This will be explained in more detail in section \ref{secsigndiff}.

\begin{table}
\begin{center} 
\begin{tabular}{|c|c|cc|}
\hline
Vertex  & \dtt  weight & Staggered six-vertex weight & \\
\hline
7 & $-4k^2x^2 e^{2iu}\sin(\gamma_0 -u_0)\sin(\gamma_0)$ & $-\frac{1}{4}e^{2iu}\sin(\gamma_0 -u_0)\sin(\gamma_0)$ & \\
\hline
8 & $-4k^2x^2 e^{2iu}\sin(\gamma_0 -u_0)\sin(\gamma_0)$ & $-\frac{1}{4}e^{2iu}\sin(\gamma_0 -u_0)\sin(\gamma_0)$ & \\
\hline
9 & $4k^2x^2 \sin(\gamma_0 -u_0)\sin(\gamma_0)$ & $-\frac{1}{4}\sin(\gamma_0 -u_0)\sin(\gamma_0)$ & \textbf{*} \\
\hline
10 & $4k^2x^2 \sin(\gamma_0 -u_0)\sin(\gamma_0)$ & $-\frac{1}{4}\sin(\gamma_0 -u_0)\sin(\gamma_0)$ & \textbf{*} \\
\hline
11 & $4k^2x^2 e^{-i(\gamma-2u)}\sin(u_0)\sin(\gamma_0)$ & $\frac{1}{4}e^{-i(\gamma-2u)}\sin(u_0)\sin(\gamma_0)$ & \\
\hline
12 & $4k^2x^2 e^{-i(\gamma-2u)}\sin(u_0)\sin(\gamma_0)$ & $\frac{1}{4}e^{-i(\gamma-2u)}\sin(u_0)\sin(\gamma_0)$ & \\
\hline
13 & $4k^2x^2 e^{i\gamma}\sin(u_0)\sin(\gamma_0)$ & $-\frac{1}{4}e^{i\gamma}\sin(u_0)\sin(\gamma_0)$ & \textbf{*} \\
\hline
14 & $4k^2x^2 e^{i\gamma}\sin(u_0)\sin(\gamma_0)$ & $-\frac{1}{4}e^{i\gamma}\sin(u_0)\sin(\gamma_0)$ & \textbf{*} \\
\hline
15 & $-4k^2x^2 e^{-2iu}\sin(\gamma_0-u_0)\sin(\gamma_0)$ & $-\frac{1}{4}e^{-2iu}\sin(\gamma_0-u_0)\sin(\gamma_0)$ & \\
\hline
16 & $-4k^2x^2 e^{-2iu}\sin(\gamma_0-u_0)\sin(\gamma_0)$ & $-\frac{1}{4}e^{-2iu}\sin(\gamma_0-u_0)\sin(\gamma_0)$ & \\
\hline
17 & $4k^2x^2 e^{-i\gamma}\sin(u_0)\sin(\gamma_0)$ & $-\frac{1}{4}e^{-i\gamma}\sin(u_0)\sin(\gamma_0)$ & \textbf{*}\\
\hline
18 & $4k^2x^2 e^{-i\gamma}\sin(u_0)\sin(\gamma_0)$ & $-\frac{1}{4}e^{-i\gamma}\sin(u_0)\sin(\gamma_0)$ & \textbf{*} \\
\hline
19 & $-4k^2x^2 \sin(\gamma_0-u_0)\sin(u_0)$ & $-\frac{1}{4}\sin(\gamma_0-u_0)\sin(u_0)$ & \\
\hline
20 & $-4k^2x^2 \sin(\gamma_0-u_0)\sin(u_0)$ & $-\frac{1}{4}\sin(\gamma_0-u_0)\sin(u_0)$ & \\
\hline
21 & $-4k^2x^2 \sin(\gamma_0-u_0)\sin(u_0)$ & $-\frac{1}{4}\sin(\gamma_0-u_0)\sin(u_0)$ & \\
\hline
22 & $-4k^2x^2 \sin(\gamma_0-u_0)\sin(u_0)$ & $-\frac{1}{4}\sin(\gamma_0-u_0)\sin(u_0)$ & \\
\hline
23 & $-4k^2x^2 \sin(\gamma_0-u_0)\sin(u_0)$ & $-\frac{1}{4}\sin(\gamma_0-u_0)\sin(u_0)$ & \\
\hline
24 & $-4k^2x^2 \sin(\gamma_0-u_0)\sin(u_0)$ & $-\frac{1}{4}\sin(\gamma_0-u_0)\sin(u_0)$ & \\
\hline
25 & $-4k^2x^2 \sin(\gamma_0-u_0)\sin(u_0)$ & $-\frac{1}{4}\sin(\gamma_0-u_0)\sin(u_0)$ & \\
\hline
26 & $-4k^2x^2 \sin(\gamma_0-u_0)\sin(u_0)$ & $-\frac{1}{4}\sin(\gamma_0-u_0)\sin(u_0)$ & \\
\hline
27 & $4k^2x^2 e^{i(\gamma-2u)}\sin(u_0)\sin(\gamma_0)$ & $\frac{1}{4}e^{i(\gamma-2u)}\sin(u_0)\sin(\gamma_0)$ & \\
\hline
28 & $4k^2x^2 e^{i(\gamma-2u)}\sin(u_0)\sin(\gamma_0)$ & $\frac{1}{4}e^{i(\gamma-2u)}\sin(u_0)\sin(\gamma_0)$ & \\
\hline
29 & $4k^2x^2 \sin(\gamma_0-u_0)\sin(\gamma_0)$ & $-\frac{1}{4}\sin(\gamma_0-u_0)\sin(\gamma_0)$ & \textbf{*} \\
\hline
30 & $4k^2x^2 \sin(\gamma_0-u_0)\sin(\gamma_0)$ & $-\frac{1}{4}\sin(\gamma_0-u_0)\sin(\gamma_0)$ & \textbf{*} \\
\hline
\end{tabular}
\end{center}
\caption{Correspondence between Boltzmann weights (continued). Vertices 7 to 30.}\label{vert7to30}
\end{table}

\subsubsection{Vertices 31 to 38}\label{secvert31to38}
This section will present the calculation of the Boltzmann weights of vertices with no arrows. These vertices are labelled 31 to 38 in figure \ref{38vertices} and correspond to the terms 
$E_{\tilde{3}\tilde{2}}\otimes E_{\tilde{3}\tilde{2}}$, $E_{\tilde{2}\tilde{3}}\otimes E_{\tilde{2}\tilde{3}}$,
$E_{\tilde{2}\tilde{2}}\otimes E_{\tilde{3}\tilde{3}}$, 
$E_{\tilde{3}\tilde{3}}\otimes E_{\tilde{2}\tilde{2}}$,
$E_{\tilde{3}\tilde{2}}\otimes E_{\tilde{2}\tilde{3}}$,  
$E_{\tilde{2}\tilde{3}}\otimes E_{\tilde{3}\tilde{2}}$, 
$E_{\tilde{2}\tilde{2}}\otimes E_{\tilde{2}\tilde{2}}$,   $E_{\tilde{3}\tilde{3}}\otimes E_{\tilde{3}\tilde{3}}$. 

Consider the terms in the expansion of the \dtt $R$-matrix:
\beq\label{terms31to38}
\begin{aligned}
& E_{22}\otimes E_{22}[k(x^2-1)(x^2-k^2)-\frac{1}{2}(k^2-1)(k+1)x(x+1)(x-k)]\\
+ & E_{33}\otimes E_{33}[k(x^2-1)(x^2-k^2)-\frac{1}{2}(k^2-1)(k+1)x(x+1)(x-k)]\\
+ & E_{22}\otimes E_{33}[k(x^2-1)(x^2-k^2)+\frac{1}{2}(k^2-1)(k+1)x(x-1)(x+k)]\\
+ & E_{33}\otimes E_{22}[k(x^2-1)(x^2-k^2)+\frac{1}{2}(k^2-1)(k+1)x(x-1)(x+k)]\\
+ & E_{32}\otimes E_{23}[\frac{1}{2}(k^2-1)(k-1)x(x+1)(x+k)]\\
+ & E_{23}\otimes E_{32}[\frac{1}{2}(k^2-1)(k-1)x(x+1)(x+k)]\\
+ & E_{32}\otimes E_{32}[-\frac{1}{2}(k^2-1)(k-1)x(x-1)(x-k)]\\
+ & E_{23}\otimes E_{23}[-\frac{1}{2}(k^2-1)(k-1)x(x-1)(x-k)] \,.
\end{aligned}
\eeq
Now use the easily verified expressions:
\begin{subequations}
\begin{eqnarray}
E_{22} \otimes E_{22} + E_{33} \otimes E_{33} = & \!\!\! \frac{1}{2}[ E_{\tilde{2}\tilde{2}}\otimes E_{\tilde{2}\tilde{2}}+E_{\tilde{2}\tilde{3}}\otimes E_{\tilde{2}\tilde{3}}+E_{\tilde{3}\tilde{2}}\otimes E_{\tilde{3}\tilde{2}}+E_{\tilde{2}\tilde{3}}\otimes E_{\tilde{3}\tilde{2}}+ \nonumber \\ & E_{\tilde{3}\tilde{2}}\otimes E_{\tilde{2}\tilde{3}}+E_{\tilde{2}\tilde{2}}\otimes E_{\tilde{3}\tilde{3}}+E_{\tilde{3}\tilde{3}}\otimes E_{\tilde{2}\tilde{2}}+E_{\tilde{3}\tilde{3}}\otimes E_{\tilde{3}\tilde{3}}] \,, \\
E_{22} \otimes E_{33} + E_{33} \otimes E_{22} = & \!\!\! \frac{1}{2}[ E_{\tilde{2}\tilde{2}}\otimes E_{\tilde{2}\tilde{2}}-E_{\tilde{2}\tilde{3}}\otimes E_{\tilde{2}\tilde{3}}-E_{\tilde{3}\tilde{2}}\otimes E_{\tilde{3}\tilde{2}}-E_{\tilde{2}\tilde{3}}\otimes E_{\tilde{3}\tilde{2}}- \nonumber \\ & E_{\tilde{3}\tilde{2}}\otimes E_{\tilde{2}\tilde{3}}+E_{\tilde{2}\tilde{2}}\otimes E_{\tilde{3}\tilde{3}}+E_{\tilde{3}\tilde{3}}\otimes E_{\tilde{2}\tilde{2}}+E_{\tilde{3}\tilde{3}}\otimes E_{\tilde{3}\tilde{3}}] \,, \\
E_{32} \otimes E_{23} + E_{23} \otimes E_{32} = & \!\!\! \frac{1}{2}[ E_{\tilde{2}\tilde{2}}\otimes E_{\tilde{2}\tilde{2}}-E_{\tilde{2}\tilde{3}}\otimes E_{\tilde{2}\tilde{3}}-E_{\tilde{3}\tilde{2}}\otimes E_{\tilde{3}\tilde{2}}+E_{\tilde{2}\tilde{3}}\otimes E_{\tilde{3}\tilde{2}}+ \nonumber \\ & E_{\tilde{3}\tilde{2}}\otimes E_{\tilde{2}\tilde{3}}-E_{\tilde{2}\tilde{2}}\otimes E_{\tilde{3}\tilde{3}}-E_{\tilde{3}\tilde{3}}\otimes E_{\tilde{2}\tilde{2}}+E_{\tilde{3}\tilde{3}}\otimes E_{\tilde{3}\tilde{3}}] \,, \\
E_{32} \otimes E_{32} + E_{23} \otimes E_{23} = & \!\!\! \frac{1}{2}[ E_{\tilde{2}\tilde{2}}\otimes E_{\tilde{2}\tilde{2}}+E_{\tilde{2}\tilde{3}}\otimes E_{\tilde{2}\tilde{3}}+E_{\tilde{3}\tilde{2}}\otimes E_{\tilde{3}\tilde{2}}-E_{\tilde{2}\tilde{3}}\otimes E_{\tilde{3}\tilde{2}}- \nonumber \\ & E_{\tilde{3}\tilde{2}}\otimes E_{\tilde{2}\tilde{3}}-E_{\tilde{2}\tilde{2}}\otimes E_{\tilde{3}\tilde{3}}-E_{\tilde{3}\tilde{3}}\otimes E_{\tilde{2}\tilde{2}}+E_{\tilde{3}\tilde{3}}\otimes E_{\tilde{3}\tilde{3}}]
\end{eqnarray}
\end{subequations}
to write the terms in (\ref{terms31to38}) as
\beq
\begin{aligned}
&\frac{1}{2} \left[ k(x^2-1)(x^2-k^2)-\frac{1}{2}(k^2-1)(k+1)x(x+1)(x-k) \right] \times \\
&\qquad \big( E_{\tilde{2}\tilde{2}}\otimes E_{\tilde{2}\tilde{2}}+E_{\tilde{2}\tilde{3}}\otimes E_{\tilde{2}\tilde{3}} + E_{\tilde{3}\tilde{2}}\otimes E_{\tilde{3}\tilde{2}}+E_{\tilde{2}\tilde{3}}\otimes E_{\tilde{3}\tilde{2}} + \\
& \qquad \ \, E_{\tilde{3}\tilde{2}}\otimes E_{\tilde{2}\tilde{3}}+E_{\tilde{2}\tilde{2}}\otimes E_{\tilde{3}\tilde{3}} +  E_{\tilde{3}\tilde{3}}\otimes E_{\tilde{2}\tilde{2}}+E_{\tilde{3}\tilde{3}}\otimes E_{\tilde{3}\tilde{3}} \big) \\
+ & \frac{1}{2} \left[ k(x^2-1)(x^2-k^2)+\frac{1}{2}(k^2-1)(k+1)x(x-1)(x+k) \right] \times \\
&\qquad \big( E_{\tilde{2}\tilde{2}}\otimes E_{\tilde{2}\tilde{2}}-E_{\tilde{2}\tilde{3}}\otimes E_{\tilde{2}\tilde{3}} - E_{\tilde{3}\tilde{2}}\otimes E_{\tilde{3}\tilde{2}}-E_{\tilde{2}\tilde{3}}\otimes E_{\tilde{3}\tilde{2}} -\\
& \qquad \ \, E_{\tilde{3}\tilde{2}}\otimes E_{\tilde{2}\tilde{3}}+E_{\tilde{2}\tilde{2}}\otimes E_{\tilde{3}\tilde{3}}+
  E_{\tilde{3}\tilde{3}}\otimes E_{\tilde{2}\tilde{2}}+E_{\tilde{3}\tilde{3}}\otimes E_{\tilde{3}\tilde{3}} \big) \\
+ & \left[ \frac{1}{4}(k^2-1)(k-1)x(x+1)(x+k) \right] \times \\
&\qquad \big( E_{\tilde{2}\tilde{2}}\otimes E_{\tilde{2}\tilde{2}}-E_{\tilde{2}\tilde{3}}\otimes E_{\tilde{2}\tilde{3}}  - E_{\tilde{3}\tilde{2}}\otimes E_{\tilde{3}\tilde{2}}+E_{\tilde{2}\tilde{3}}\otimes E_{\tilde{3}\tilde{2}} + \\
& \qquad \ \, E_{\tilde{3}\tilde{2}}\otimes E_{\tilde{2}\tilde{3}}-E_{\tilde{2}\tilde{2}}\otimes E_{\tilde{3}\tilde{3}} - E_{\tilde{3}\tilde{3}}\otimes E_{\tilde{2}\tilde{2}}+E_{\tilde{3}\tilde{3}}\otimes E_{\tilde{3}\tilde{3}} \big) \\
- & \left[ \frac{1}{4}(k^2-1)(k-1)x(x-1)(x-k) \right] \times \\
&\qquad \big( E_{\tilde{2}\tilde{2}}\otimes E_{\tilde{2}\tilde{2}}+E_{\tilde{2}\tilde{3}}\otimes E_{\tilde{2}\tilde{3}} + E_{\tilde{3}\tilde{2}}\otimes E_{\tilde{3}\tilde{2}}-E_{\tilde{2}\tilde{3}}\otimes E_{\tilde{3}\tilde{2}} - \\
&\qquad \ \, E_{\tilde{3}\tilde{2}}\otimes E_{\tilde{2}\tilde{3}}-E_{\tilde{2}\tilde{2}}\otimes E_{\tilde{3}\tilde{3}} - E_{\tilde{3}\tilde{3}}\otimes E_{\tilde{2}\tilde{2}}+E_{\tilde{3}\tilde{3}}\otimes E_{\tilde{3}\tilde{3}} \big) \,.
\end{aligned}
\eeq
We now collect coefficients of each of the terms. The coefficient of, for example, $E_{\tilde{2}\tilde{2}}\otimes E_{\tilde{2}\tilde{2}}$ reduces to $k(x^2-1)(x^2-k^2)+x^2(k^2-1)^2$ which after applying \eqref{kx_corr} becomes $-4k^2 x^2[\sin^2(\gamma_0)-\sin(\gamma_0-u_0)\sin(u_0)]$, which is exactly the coefficient of vertex 37 in figure \ref{38vertices} in both the \dtt model and the staggered six-vertex model. The results for the other coefficients are summarised in table \ref{vert31to38}.

\begin{table}
\begin{center} 
\begin{tabular}{|c|c|cc|}
\hline
Vertex  & \dtt weight & Staggered six-vertex weight & \\
\hline
31 & $4k^2x^2\sin(u_0)\sin(\gamma_0)$ & $-\frac{1}{4}\sin(u_0)\sin(\gamma_0)$ & \textbf{*} \\
\hline
32 & $4k^2x^2\sin(u_0)\sin(\gamma_0)$ & $-\frac{1}{4}\sin(u_0)\sin(\gamma_0)$ & \textbf{*} \\
\hline
33 & $-4k^2x^2\sin(\gamma_0-u_0)\sin(u_0)$ & $-\frac{1}{4}\sin(\gamma_0-u_0)\sin(u_0)$ & \\
\hline
34 & $-4k^2x^2\sin(\gamma_0-u_0)\sin(u_0)$ & $-\frac{1}{4}\sin(\gamma_0-u_0)\sin(u_0)$ & \\
\hline
35 & $4k^2x^2\sin(\gamma_0-u_0)\sin(\gamma_0)$ &  $-\frac{1}{4}\sin(\gamma_0-u_0)\sin(\gamma_0)$ & \textbf{*} \\
\hline
36 & $4k^2x^2\sin(\gamma_0-u_0)\sin(\gamma_0)$ &  $-\frac{1}{4}\sin(\gamma_0-u_0)\sin(\gamma_0)$ & \textbf{*} \\
\hline
37 & $-4k^2x^2(\sin^2(\gamma_0)+\sin(\gamma_0-u_0)\sin(u_0))$ & $-\frac{1}{4}(\sin^2(\gamma_0)+\sin(\gamma_0-u_0)\sin(u_0))$ & \\
\hline
38 & $-4k^2x^2(\sin^2(\gamma_0)+\sin(\gamma_0-u_0)\sin(u_0))$ & $-\frac{1}{4}(\sin^2(\gamma_0)+\sin(\gamma_0-u_0)\sin(u_0))$ & \\
\hline
\end{tabular}
\end{center}
\caption{Correspondence between Boltzmann weights (continued). Vertices 31 to 38.} \label{vert31to38}
\end{table}

\subsubsection{Sign differences}\label{secsigndiff}
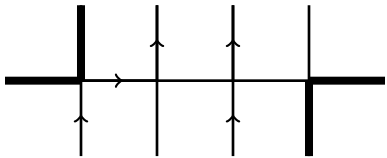
\begin{figure}
	\centering
	\begin{tikzpicture}
		
\begin{scope}[thick,decoration={
    markings,
    mark=at position 0.55 with {\arrow{>}}}
    ] 
\draw[black,line width = 1pt,postaction={decorate}] (1,0)--(1,1);
\end{scope}	

\begin{scope}[thick,decoration={
    markings,
    mark=at position 0.55 with {\arrow{>}}}
    ] 
\draw[black,line width = 1pt,postaction={decorate}] (3,0)--(3,1);
\end{scope}	

\begin{scope}[thick,decoration={
    markings,
    mark=at position 0.55 with {\arrow{>}}}
    ] 
\draw[black,line width = 1pt,postaction={decorate}] (1,1)--(2,1);
\end{scope}	

\begin{scope}[thick,decoration={
    markings,
    mark=at position 0.55 with {\arrow{>}}}
    ] 
\draw[black,line width = 1pt,postaction={decorate}] (2,1)--(2,2);
\end{scope}	

\begin{scope}[thick,decoration={
    markings,
    mark=at position 0.55 with {\arrow{>}}}
    ] 
\draw[black,line width = 1pt,postaction={decorate}] (3,1)--(3,2);
\end{scope}	
	
\draw[black,line width = 3pt](0,1)--(1,1);	
\draw[black,line width = 3pt](1,1)--(1,2);
\draw[black,line width = 3pt](4,1)--(5,1);
\draw[black,line width = 3pt](4,0)--(4,1);

\draw[black,line width = 1pt](1,1)--(5,1);

\draw[black,line width = 1pt](1,1)--(1,2);
\draw[black,line width = 1pt](2,0)--(2,2);
\draw[black,line width = 1pt](3,1)--(3,2);
\draw[black,line width = 1pt](4,0)--(4,2);

\end{tikzpicture}
\caption{A configuration of vertices in one row generated by the action of the transfer matrix. The vertices on the far left and far right of the figure correspond to vertices (9) and (35) in figure \ref{38vertices}. The Boltzmann weights of these two vertices can be found in tables \ref{vert1to6} and \ref{vert31to38} respectively where we observe that both of them have a \textbf{*} beside them, meaning that their signs are opposite to the signs of the corresponding vertices of the staggered six vertex model. The two minus signs cancel each other out. More generally, the periodic boundary conditions ensure that there are will always be an even number of vertices with opposite signs in the two models.}\label{tmatperiodic}
\end{figure}

As was briefly touched upon, there are some Boltzmann weights in the \dtt construction of the model that differ by a sign from the Boltzmann weights in the staggered six-vertex version of the model. These vertices have been highlighted by asterisks in the right columns of tables \ref{vert1to6}--\ref{vert31to38}. From figure \ref{38vertices} we observe that all of these vertices are such that there is one horizontal thick line and one vertical thick line. Observe then that, as well as the conservation of the direction of arrows, all vertices conserve the parity of the number of thick lines. In particular, if there is one incoming thick line there must be one outgoing thick line, and if there are two incoming thick lines there must be \textit{either} two outgoing thick lines or no outgoing thick lines. A consequence of this is that, when periodic boundary conditions are imposed, any given configuration of vertices generated by a transfer matrix must contain an even number of these vertices with asterisks, and hence the minus signs all cancel. This is highlighted in figure \ref{tmatperiodic}.

Figure \ref{tmatperiodic} resolves the sign problem when studying the model with periodic boundary conditions. With open boundary conditions, however, it is not so clear that the transfer matrices of the two models will be equal since, for a general open boundary condition, we are allowed to have odd numbers of vertices which differ by a sign in the two models. It will turn out nonetheless that the boundary conditions we are interested in also preserve the parity of the number of thick lines and the transfer matrix will ensure that we again only encounter configurations with an even number of these vertices with asterisks. This preservation of the parity of thick and thin lines turns out to be a result of a symmetry under a lattice operator denoted by $C$, which was first introduced in \cite{ikhlef2008}.  This operator is most conveniently expressed as
\beq\label{Cdef}
C=C_1C_3 \cdots C_{2L-1}
\eeq
where
\beq\label{cdef}
C_i =1-\frac{1}{\cos\gamma}e_i \,,
\eeq
and $e_i$ is a generator of the Temperley-Lieb (TL) algebra \cite{TL71}. For a system defined on $N$ sites, the TL algebra is defined in terms of
generators $e_i$ with $i=1,2,\ldots,N-1$, subject to the relations
\begin{subequations}
\label{TLrelations}
\begin{eqnarray}
e_i^2 &=&  \sqrt{Q} e_i \,, \\
e_i e_{i\pm1}e_i &=& e_i \,, \\
e_i e_j &=& e_je_i \text{ for } |i-j|\geq2 \,.
\end{eqnarray}
\end{subequations}
Both the $C$ operator and the TL algebra will play important roles in what follows and we shall discuss them more fully below.

\section{The open \dtt model}\label{secopengen}
\begin{figure}
	\centering
	\begin{tikzpicture}
		
\begin{scope}[thick,decoration={
    markings,
    mark=at position 0.55 with {\arrow{>}}}
    ] 
\draw[black,line width = 1pt,postaction={decorate}] (1,0)--(1,1);
\end{scope}	

\begin{scope}[thick,decoration={
    markings,
    mark=at position 0.55 with {\arrow{>}}}
    ] 
\draw[black,line width = 1pt,postaction={decorate}] (3,0)--(3,1);
\end{scope}

\begin{scope}[thick,decoration={
    markings,
    mark=at position 0.55 with {\arrow{>}}}
    ] 
\draw[black,line width = 1pt,postaction={decorate}] (1,1)--(2,1);
\end{scope}		

\begin{scope}[thick,decoration={
    markings,
    mark=at position 0.55 with {\arrow{>}}}
    ] 
\draw[black,line width = 1pt,postaction={decorate}] (2,1)--(2,2);
\end{scope}	

\begin{scope}[thick,decoration={
    markings,
    mark=at position 0.55 with {\arrow{>}}}
    ] 
\draw[black,line width = 1pt,postaction={decorate}] (2,2)--(2,3);
\end{scope}	

\begin{scope}[thick,decoration={
    markings,
    mark=at position 0.55 with {\arrow{>}}}
    ] 
\draw[black,line width = 1pt,postaction={decorate}] (3,1)--(4,1);
\end{scope}	

\begin{scope}[thick,decoration={
    markings,
    mark=at position 0.55 with {\arrow{>}}}
    ] 
\draw[black,line width = 1pt,postaction={decorate}] (4,1)--(4,2);
\end{scope}	

\begin{scope}[thick,decoration={
    markings,
    mark=at position 0.55 with {\arrow{>}}}
    ] 
\draw[black,line width = 1pt,postaction={decorate}] (4,2)--(4,3);
\end{scope}	
	

\draw[black,line width = 3pt] (1,1)-- (0,1) .. controls (-0.5,1.5) .. (0,2)--(1,2);
\draw[black,line width = 3pt] (1,1)-- (0,1) .. controls (-0.5,1.5) .. (0,2)--(1,2);

\draw[black,line width = 1pt](4,1)--(5,1) .. controls (5.5,1.5) .. (5,2)--(4,2);

\draw[black,line width = 3pt](2,0)--(2,1);
\draw[black,line width = 3pt](3,1)--(3,2);
\draw[black,line width = 3pt](2,1)--(3,1);
\draw[black,line width = 3pt](3,2)--(3,3);

\draw[black,line width = 3pt](1,1)--(1,2);
\draw[black,line width = 1pt](1,2)--(4,2);
	
\draw[black,line width = 1pt](1,1)--(1,3);
\draw[black,line width = 1pt](2,0)--(2,3);
\draw[black,line width = 1pt](3,1)--(3,3);
\draw[black,line width = 1pt](4,0)--(4,3);

\end{tikzpicture}
\caption{Two rows of vertices constructed from the transfer matrix. The fact that $K$ becomes diagonal in the appropriate basis implies that the parity of the number of thick lines are conserved. This ensures that there are an even number of vertices that differ by a sign in the \dtt model and the staggered six-vertex model.}\label{tmatopen}
\end{figure}
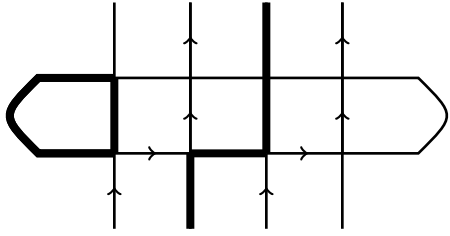

\no To construct a closed integrable model we start with an $R$-matrix that acts on the space $V\otimes V$, where $V$ is a $d$-dimensional space, and satisfies the Yang-Baxter equation. We then define a transfer matrix in the following way:

\beq\label{tmatrixperiodic}
t(u)=\Tr_a(R_{a1}(u)...R_{aL}(u)) \,,
\eeq
where the trace is over the ``auxiliary space'' $a$. To construct an integrable model with open boundary conditions, however, in addition to the $R$-matrix that satisfies the Yang-Baxter equation we need to consider a particular $d\times d$ matrix acting at the boundary: the $K$-matrix. As a matter of fact, we shall need a $K$-matrix for both the left and right boundaries which we will denote as $K^-$ and $K^+$, respectively. We require that $K^-(u)$ satisfy an analogue of the Yang-Baxter equation, the so-called boundary Yang-Baxter equation \cite{sklyanin1988}
\beq\label{reflection}
R_{12}(u-v)K_1^-(u)R_{21}(u+v)K_2^-(v)=K_2^-(u)R_{12}(u+v)K_1^-(u)R_{21}(u-v) \,,
\eeq
so that the two-row transfer matrix (cf.\ figure \ref{tmatopen})
\beq\label{tmatrixopen}
t(u)=\Tr_a K_a^+(u)R_{a1}(u)...R_{aL}(u)K_a^-(u)R_{1a}(u)...R_{La}(u)
\eeq
will be integrable, i.e., satisfy $[t(u),t(v)]=0$ for all $u,v$. To ensure that the right $K$-matrix, $K^+(u)$, satisfies the analogue of (\ref{reflection}) on the right boundary we take
\beq\label{Kright}
K^+(\lambda)=K^{-{\rm t}}(-\rho-\lambda)M \,,
\eeq
where $\rho$ and $M$ are model dependent and ${\rm t}$ denotes an antiautomorphism which coincicdes here with the usual matrix transposition. In the case that we are considering here, i.e., the \dtt model, we have $\rho=-\log k$ and $M=\text{diag}(k,1,1,k^{-1})$ \cite{martinsd22}. Here we will consider a particular $K$-matrix that satisfies (\ref{reflection}) \cite{nepomechied22}:

\beq\label{kmat1}
K^-(\lambda)=\begin{bmatrix}
Y_1(\lambda) & 0 & 0 & 0\\
0 & Y_2(\lambda) & Y_5(\lambda) & 0\\
0 & Y_6(\lambda) & Y_3(\lambda)& 0\\
0 & 0 & 0 &  Y_4(\lambda)
\end{bmatrix} \,,
\eeq
with 
\begin{subequations}
\label{kmat2}
\begin{eqnarray}
Y_1(\lambda) &=& -e^{-\lambda}(e^{2\lambda}+k) \,, \\
Y_4(\lambda) &=& -e^{3\lambda}(e^{2\lambda}+k) \,, \\
Y_2(\lambda) &=&Y_3(\lambda)=-\frac{1}{2}(1+e^{2\lambda})e^{\lambda}(1+k) \,, \\
Y_5(\lambda) &=& Y_6(\lambda)=\frac{1}{2}(e^{2\lambda}-1)(1-k)e^{\lambda} \,,
\end{eqnarray}
\end{subequations}
and we recall that $u$ and $k$ satisfy the relations \eqref{kx_corr}.
Recall now the discussion in section \ref{secsigndiff} about the particular Boltzmann weights in tables \ref{vert1to6}--\ref{vert31to38} that differed by a sign when considering the \dtt model and the staggered six-vertex model. This issue was resolved by noticing that, when periodic boundary conditions are imposed, there is always an even number of these vertices and hence the the transfer matrix built from either $R$-matrix is the same.

Now that we are considering open boundary conditions we can no longer rely on the same argument. Notice however, that in the basis defined in equation (\ref{cbasis}) the $K$-matrix in equation (\ref{kmat1}) becomes diagonal:

\beq
K^-(\lambda)\rightarrow\begin{bmatrix}
Y_1(\lambda) & 0 & 0 & 0\\
0 & Y_2(\lambda)+Y_5(\lambda) & 0 & 0\\
0 & 0 & Y_2(\lambda)-Y_6(\lambda)& 0\\
0 & 0 & 0 &  Y_4(\lambda)
\end{bmatrix}
\eeq 
The $K$-matrix being diagonal ensures that we have conservation of both thick and thin lines at the boundary and that in any given configuration we will again have an even number of vertices that differ by a sign in the two models. See figure \ref{tmatopen} for an illustration.

The fact that the $K$-matrix is diagonal in this basis comes from the fact that it commutes with the $C$-operator defined in equation (\ref{cdef}). The basis in (\ref{cbasis}) was in fact chosen since each of the basis vectors are eigenvectors of the $C$ operator. The $K$-matrix then satisfies
\beq
[K,C]=0 \,.
\eeq
This symmetry will be discussed in more detail in section \ref{addsym} and it will turn out to account for the extra degeneracies observed in the spectrum of the open \dtt transfer matrix/Hamiltonian.

\subsection{Hamiltonian limit}
Following the construction in \cite{sklyanin1988} we can define an open integrable Hamiltonian from an open integrable transfer matrix in the following way:
\beq\label{tmatham}
t'(0)=2H \Tr K^+(0)+\Tr K^{+'}(0) \,,
\eeq
which gives
\beq\label{hamintopen}
H=\sum\limits_{n=1}^{L-1}H_{n,n+1}+\frac{1}{2}K^{-'}_1(0)+\frac{\Tr_0K^+_0(0)H_{L0}}{\Tr K^+_0(0)} \,,
\eeq
where $H_{n,n+1}=P_{n,n+1}\frac{{\rm d}}{{\rm d}\lambda}R_{n,n+1}(\lambda)|_{\lambda=0}$; the subscripts indicate on which tensor there is a non-trivial action. Here, $P$ denotes the permutation operator. Recall that the transfer matrix for the periodic case is given by (\ref{tmatrixperiodic}) and the corresponding Hamiltonian is again obtained by taking the derivative with respect to the spectral parameter. Up to overall normalisation terms, the periodic Hamiltonian can be written \cite{ikhlef2008,ikhlef2009}:
\beq\label{hperiodic}
\mathcal{H}=2\cos\gamma\sum\limits_{m=1}^{2L-1}e_m-(e_me_{m+1}+e_{m+1}e_m) \,,
\eeq
where the TL generators $e_m$ satisfy \eqref{TLrelations}.
The open Hamiltonian in (\ref{hamintopen}) can be similarly written as
\beq\label{hopen}
\mathcal{H}=A_{\rm left}+A_{\rm right}+\cos\gamma(e_1+e_{2L-1})+2\cos\gamma\sum\limits_{m=2}^{2L-2}e_m-\sum\limits_{m=1}^{2L-2}(e_me_{m+1}+e_{m+1}e_m) \,,
\eeq
where $A_{\rm left}$ and $A_{\rm right}$ are the boundary terms arising from the second and third terms in equation (\ref{hamintopen}) and can be written as
\beq\label{lbdry}
A_{\rm left}=
\begin{pmatrix}
	i\sin2\gamma & 0 & 0 & 0 \\
	0 & -\frac{\sin^2\gamma}{\cos\gamma}e^{i\gamma} & \frac{\sin^2\gamma}{\cos\gamma} & 0\\
	0 & \frac{\sin^2\gamma}{\cos\gamma} & -\frac{\sin^2\gamma}{\cos\gamma}e^{-i\gamma} & 0\\
	0 & 0 & 0 & -i\sin2\gamma
\end{pmatrix}
\otimes
\mathbb{I}^{\otimes 2L-2}
\eeq
and
\beq\label{lbdry}
A_{\rm right}=
\mathbb{I}^{\otimes 2L-2}
\otimes
\begin{pmatrix}
	-i\sin2\gamma & 0 & 0 & 0 \\
	0 & -\frac{\sin^2\gamma}{\cos\gamma}e^{i\gamma} & \frac{\sin^2\gamma}{\cos\gamma} & 0\\
	0 & \frac{\sin^2\gamma}{\cos\gamma} & -\frac{\sin^2\gamma}{\cos\gamma}e^{-i\gamma} & 0\\
	0 & 0 & 0 & i\sin2\gamma
\end{pmatrix}
\eeq
after subtracting terms proportional to the identity. The usual representation of the TL generators $e_i$ in the vertex model are given by
\beq\label{ei}
e_i=
\mathbb{I}^{\otimes i-1}
\otimes
\begin{pmatrix}
	0 & 0 & 0 & 0 \\
	0 & e^{-i\gamma} & 1 & 0\\
	0 & 1 & e^{i\gamma} & 0\\
	0 & 0 & 0 & 0 
\end{pmatrix}
\otimes
\mathbb{I}^{\otimes 2L-i-1} \,,
\eeq
but we shall need as well another representation of the TL algebra
\beq\label{eitilde}
\tilde{e}_i=
\mathbb{I}^{\otimes i-1}
\otimes
\begin{pmatrix}
	0 & 0 & 0 & 0 \\
	0 & e^{i\gamma} & -1 & 0\\
	0 & -1 & e^{-i\gamma} & 0\\
	0 & 0 & 0 & 0 
\end{pmatrix}
\otimes
\mathbb{I}^{\otimes 2L-i-1} \,,
\eeq
which can also be checked to satisfy \eqref{TLrelations}.
We can now write 
\beq\label{tlleft}
A_{\rm left} = -\frac{\sin^2 \gamma}{\cos\gamma}\tilde{e}_1+i\sin2\gamma \left( \frac{1}{2}\sigma_1^z+\frac{1}{2}\sigma_2^z \right)
\eeq
and
\beq\label{tlright}
A_{\rm right} = -\frac{\sin^2 \gamma}{\cos\gamma}\tilde{e}_{2L-1}-i\sin2\gamma \left( \frac{1}{2}\sigma_1^z+\frac{1}{2}\sigma_2^z \right) \,.
\eeq
By expanding the TL generators $e_i$ and $\tilde{e}_i$ in terms of Pauli matrices,
\begin{subequations}
\label{esigma}
\begin{eqnarray}
e_i &=& \frac{1}{2}\left[\sigma_i^x\sigma_{i+1}^x+\sigma_i^y\sigma_{i+1}^y-\cos\gamma\sigma_i^z\sigma_{i+1}^z+\cos\gamma-i\sin\gamma(\sigma_i^z-\sigma_{i+1}^z) \right] \,, \\
\tilde{e}_i &=& \frac{1}{2}\left[-\sigma_i^x\sigma_{i+1}^x-\sigma_i^y\sigma_{i+1}^y-\cos\gamma\sigma_i^z\sigma_{i+1}^z+\cos\gamma+i\sin\gamma(\sigma_i^z-\sigma_{i+1}^z) \right] \,,
\end{eqnarray}
\end{subequations}
we can see that the Hamiltonian (\ref{hopen}) can be written entirely in terms of $\tilde{e}_i$ instead of $e_i$. The additional Pauli matrices in equations (\ref{tlleft}) and (\ref{tlright}) disappear, and we get
\beq\label{hopen2}
\mathcal{H}= \left( \cos\gamma-\frac{\sin^2 \gamma}{\cos\gamma} \right) (\tilde{e}_1+\tilde{e}_{2L-1})+2\cos\gamma\sum\limits_{m=2}^{2L-2}\tilde{e}_m-\sum\limits_{m=1}^{2L-2}(\tilde{e}_m\tilde{e}_{m+1}+\tilde{e}_{m+1}\tilde{e}_m) \,,
\eeq
which can be rewritten as
\beq\label{hopen3}
\mathcal{H}=-\frac{1}{\cos\gamma}(\tilde{e}_1+\tilde{e}_{2L-1})+2\cos\gamma\sum\limits_{m=1}^{2L-1}\tilde{e}_m-\sum\limits_{m=1}^{2L-2}(\tilde{e}_m\tilde{e}_{m+1}+\tilde{e}_{m+1}\tilde{e}_m) \,.
\eeq
Evidently, we can swap all of the $\tilde{e}_i\rightarrow e_i$ without changing the spectrum. Hence we get finally
\beq\label{hopen4}
\mathcal{H}=-\frac{1}{\cos\gamma}(e_1+e_{2L-1})+2\cos\gamma\sum\limits_{m=1}^{2L-1}e_m-\sum\limits_{m=1}^{2L-2}(e_me_{m+1}+e_{m+1}e_m) \,.
\eeq
Since the Hamiltonian in (\ref{hopen4}) arises from a Hamiltonian of the form (\ref{hamintopen}), it is integrable and solvable by Bethe Ansatz. We present its Bethe Ansatz solution in section \ref{d22type2}.

\subsection{Additional symmetries}\label{addsym}
It was found in \cite{martinsd22} and \cite{nepomechied22} that the transfer matrix (\ref{tmatrixopen})---or, equivalently, the Hamiltonian (\ref{hamintopen})---has a particular pattern of degeneracies that suggests the open chain is invariant under the action of generators of some quantum group. The observed symmetry is very similar to what one would expect if the chain were invariant under the action of the $U_q(sl(2))$ quantum group, but in fact we have even more degeneracies than would be expected if the full symmetry group was $U_q(sl(2))$.

Consider the degeneracies of the \dtt chain in table \ref{d22degens} compared with those of the expected degeneracies from a chain with just $U_q(sl(2))$ symmetry. Let us first explain the notation. On the $U_q(sl(2))$ side, $[j]$ denotes the spin-$(j-1)/2$ representation dimension $j$, and more generally $[j]$ refers to a $j$-dimensional representation of the corresponding symmetry. A decomposition like $2[1]\oplus3[3]\oplus[5]$, for example, means that there are two eigenvalues with degeneracy 1, three with degeneracy 3 and one with degeneracy 5, corresponding to a total dimension of $2 \times 1 + 3 \times 3 + 1 \times 5 = 16$.

At size $L=2$ we see that two of the 3 times degenerate eigenvalues in the $U_q(sl(2))$ chain ``become'' a 6 times degenerate eigenvalue in the \dtt chain; the symmetry group of the \dtt chain is higher. At this point it is useful to recall that a \dtt chain of length $L$ means that there are $N=2L$ sites with spin $\frac{1}{2}$, since $L$ is the number of ``Potts spins'' in one row of the classical Potts model defined by the Hamiltonian (\ref{classicalpotts}). The \dtt chain of length $L$ therefore has a Hilbert space of dimension $2^{2L}$.

We can understand the extra symmetries by studying the limit of the \dtt chain when $\gamma \rightarrow 0$, where it will be shown in section \ref{secgammazero} that the chain becomes that of two decoupled open XXX chains, and that the extra symmetry comes from the permutation of these two chains. The symmetry for finite $\gamma$ will be discussed in section \ref{secnonzerogamma}.

\begin{table}
\begin{center}
	\renewcommand{\arraystretch}{1.5} 
	\begin{tabular}{c|c|c}
		 $L$   &  $U_q(sl(2))$ & \dtt  \\
			\hline
		$2$ & $2[1]\oplus3[3]\oplus[5]$ & $2[1]\oplus[3]\oplus[5]\oplus[6]$  \\
		\hline
		$3$ & $5[1]\oplus9[3]\oplus5[5]\oplus[7]$  & $3[1]\oplus[2]\oplus3[3]\oplus[5]\oplus3[6]\oplus[7]\oplus2[10]$ \\
		\hline
		$4$ & $14[1]\oplus28[3]\oplus20[5]\oplus7[7]\oplus[9]$ & $6[1]\oplus4[3]\oplus4[2]\oplus4[5]\oplus12[6]\oplus1[7]\oplus[9]\oplus8[10]\oplus3[14]$ \\
		\end{tabular}
		\caption{The degeneracies of the \dtt spin chain of length $N=2L$ compared with those of the $U_q(sl(2))$ chain.}\label{d22degens}
	\end{center}
	\end{table}

\subsection{The $\gamma\rightarrow 0$ limit}\label{secgammazero}
Consider the Hamiltonian in (\ref{hopen4}) in the limit $\gamma\rightarrow 0$:
\beq\label{hopen5}
\mathcal{H}=(e_1+e_{2L-1})+2\sum\limits_{m=2}^{2L-2}e_m-\sum\limits_{m=1}^{2L-2}(e_me_{m+1}+e_{m+1}e_m) \,.
\eeq
Using the expression in (\ref{esigma}), this becomes
\beq\label{hopenlimit}
\mathcal{H}=-\frac{1}{2}\sum\limits_{i}^{L}(\sigma_{2i-1}^x\sigma_{2i+1}^x+\sigma_{2i-1}^y\sigma_{2i+1}^y+\sigma_{2i-1}^z\sigma_{2i+1}^z)-\frac{1}{2}\sum\limits_{i}^{L}(\sigma_{2i}^x\sigma_{2i+2}^x+\sigma_{2i}^y\sigma_{2i+2}^y+\sigma_{2i}^z\sigma_{2i+2}^z)
\eeq
up to terms proportional to the identity. The Hamiltonian in (\ref{hopenlimit}) is the sum of two decoupled open XXX chains of length $L$. (A similar observation was made for the periodic model in \cite{ikhlef2008}.)
Note that for the XXX chain, a chain of length $L$ means that the Hamiltonian acts on $L$ spin $\frac{1}{2}$ sites, unlike the \dtt chain where a chain of length $L$ means that the Hamiltonian acts on $2L$ spin-$\frac{1}{2}$ sites. This is most easily understood when considering equation (\ref{hopenlimit}), where we observed that the \dtt chain becomes equivalent to two XXX chains.

Consider first the case $L=2$. The $sl(2)$ symmetry of each individual XXX Hamiltonian is such that there are two eigenvalues, one non-degenerate and one three times degenerate. The eigenvectors of each Hamiltonian are the so-called singlet and triplet states which we will denote by $|1\rangle$ and $|3\rangle$ respectively. We will denote the corresponding eigenvalues by $\lambda_1$ and $\lambda_3$. Now consider the Hamiltonian obtained by summing the two XXX Hamiltonians. The situation is summarised in table \ref{xxxL2}. There are clearly three distinct eigenvalues given by $2\lambda_1$, $2\lambda_3$ and $\lambda_1+\lambda_3$ with the degeneracies $1$, $9$ and $6$ respectively. The eigenvectors of the full Hamiltonian are the tensor products of the eigenvectors of the two individual XXX Hamiltonians. The eigenspace of dimension $9$ comes about from the tensor product of the two spaces of dimension $3$. We can decompose this tensor product into a direct sum of spaces $|5\rangle$, $|3\rangle$ and $|1\rangle$. Note that this is just the usual tensor product of two spin-$1$ spaces into the spaces with spin $2$, $1$ and $0$. The eigenspace with dimension $6$ is more subtle. The eigenvalue $\lambda_1+\lambda_3$ corresponds to placing the eigenvector $|1\rangle$ on one XXX chain and the eigenvector $|3\rangle$ on the other. Clearly, we can swap the two chains to obtain another eigenvector with the same eigenvalue. This results then in an eigenspace of dimension $6$.

\begin{table}
\begin{center}
	\renewcommand{\arraystretch}{1.5} 
	\begin{tabular}{c|c|c|c}
		    Eigenvalue &  Eigenspace & Decomposition & Degeneracy  \\
			\hline
		$2\lambda_1$ & $|1\rangle\otimes\ |1\rangle$ & $|1\rangle$ & $1$  \\
			\hline
		$2\lambda_3$ & $|3\rangle\otimes\ |3\rangle$ & $|5\rangle\oplus\ |3\rangle\oplus\ |1\rangle$ & $9$  \\
			\hline
		$\lambda_1+\lambda_3$ & $|1\rangle\otimes\ |3\rangle\oplus\ |3\rangle\otimes\ |1\rangle$ & $|6\rangle$ & $6$  \\
		\end{tabular}
		\caption{Analysis of the spectrum of the \dtt chain for $L=2$, in the limit $\gamma \to 0$.}\label{xxxL2}
	\end{center}
	\end{table}

\begin{table}
\begin{center}
	\renewcommand{\arraystretch}{1.5} 
	\begin{tabular}{c|c|c|c}
		    Eigenvalue &  Eigenspace & Decomposition & Degeneracy  \\
			\hline
		$2\lambda_1^a$ & $|1^a\rangle\otimes\ |1^a\rangle$ & $|1\rangle$ & $1$  \\
			\hline
		$\lambda_1^a+\lambda_1^b$ & $|1^a\rangle\otimes\ |1^b\rangle\oplus\ |1^b\rangle\otimes\ |1^a\rangle$ & $|2\rangle$ & $2$  \\
			\hline
		$\lambda_1^a+\lambda_3^a$ & $|1^a\rangle\otimes\ |3^a\rangle\oplus\ |3^a\rangle\otimes\ |1^a\rangle$ & $|6\rangle$ & $6$  \\
			\hline
		$\lambda_1^a+\lambda_3^b$ & $|1^a\rangle\otimes\ |3^b\rangle\oplus\ |3^b\rangle\otimes\ |1^a\rangle$ & $|6\rangle$ & $6$  \\
			\hline
		$\lambda_1^a+\lambda_3^c$ & $|1^a\rangle\otimes\ |3^c\rangle\oplus\ |3^c\rangle\otimes\ |1^a\rangle$ & $|6\rangle$ & $6$  \\
			\hline
		$\lambda_1^a+\lambda_5$ & $|1^a\rangle\otimes\ |5\rangle\oplus\ |5\rangle\otimes\ |1^a\rangle$ & $|10\rangle$ & $5$  \\	
			\hline
		$2\lambda_1^b$ & $|1^b\rangle\otimes\ |1^b\rangle$ & $|1\rangle$ & $1$  \\
			\hline
		$\lambda_1^b+\lambda_3^a$ & $|1^b\rangle\otimes\ |3^a\rangle\oplus\ |3^a\rangle\otimes\ |1^b\rangle$ & $|6\rangle$ & $6$  \\
			\hline
		$\lambda_1^b+\lambda_3^b$ & $|1^b\rangle\otimes\ |3^b\rangle\oplus\ |3^b\rangle\otimes\ |1^b\rangle$ & $|6\rangle$ & $6$  \\
			\hline
		$\lambda_1^b+\lambda_3^c$ & $|1^b\rangle\otimes\ |3^c\rangle\oplus\ |3^c\rangle\otimes\ |1^b\rangle$ & $|6\rangle$ & $6$  \\
			\hline
		$\lambda_1^b+\lambda_5$ & $|1^b\rangle\otimes\ |5\rangle\oplus\ |5\rangle\otimes\ |1^b\rangle$ & $|10\rangle$ & $5$  \\
			\hline
		$2\lambda_3^a$ & $|3^a\rangle\otimes\ |3^a\rangle$ & $|5\rangle\oplus\ |3\rangle\oplus\ |1\rangle$ & $9$  \\
			\hline
		$\lambda_3^a+\lambda_3^b$ & $|3^a\rangle\otimes\ |3^b\rangle\oplus\ |3^b\rangle\otimes\ |3^a\rangle$ & $|10\rangle\oplus\ |6\rangle\oplus\ |2\rangle $ & $18$  \\
			\hline
		$\lambda_3^a+\lambda_3^c$ & $|3^a\rangle\otimes\ |3^c\rangle\oplus\ |3^c\rangle\otimes\ |3^a\rangle$ & $|10\rangle\oplus\ |6\rangle\oplus\ |2\rangle $ & $18$  \\	
			\hline
		$\lambda_3^a+\lambda_5$ & $|3^a\rangle\otimes\ |5\rangle\oplus\ |5\rangle\otimes\ |3^a\rangle$ & $|14\rangle\oplus\ |10\rangle\oplus\ |6\rangle $ & $30$  \\
			\hline
		$2\lambda_3^b$ & $|3^b\rangle\otimes\ |3^b\rangle$ & $|5\rangle\oplus\ |3\rangle\oplus\ |1\rangle$ & $9$  \\
			\hline
		$\lambda_3^b+\lambda_3^c$ & $|3^b\rangle\otimes\ |3^c\rangle\oplus\ |3^c\rangle\otimes\ |3^b\rangle$ & $|10\rangle\oplus\ |6\rangle\oplus\ |2\rangle $ & $18$  \\
			\hline
		$\lambda_3^b+\lambda_5$ & $|3^b\rangle\otimes\ |5\rangle\oplus\ |5\rangle\otimes\ |3^b\rangle$ & $|14\rangle\oplus\ |10\rangle\oplus\ |6\rangle $ & $30$  \\
			\hline
		$2\lambda_3^c$ & $|3^c\rangle\otimes\ |3^c\rangle$ & $|5\rangle\oplus\ |3\rangle\oplus\ |1\rangle$ & $9$  \\
			\hline
		$\lambda_3^c+\lambda_5$ & $|3^c\rangle\otimes\ |5\rangle\oplus\ |5\rangle\otimes\ |3^c\rangle$ & $|14\rangle\oplus\ |10\rangle\oplus\ |6\rangle $ & $30$  \\	
		\hline
	$2\lambda_5$ & $|5\rangle\otimes\ |5\rangle$ & $|9\rangle\oplus\ |7\rangle\oplus\ |5\rangle\oplus\ |3\rangle\oplus\ |1\rangle$ & $25$  \\
		\end{tabular}
		\caption{Spectrum of the \dtt chain with $L=4$, in the limit $\gamma \to 0$.}\label{xxxL4}
	\end{center}
	\end{table}

\subsection{Non-zero $\gamma$}\label{secnonzerogamma}

When $\gamma\neq 0$, the $sl(2)$ symmetry is replaced  by  $U_q(sl(2))$. The permutation symmetry does not hold any longer, but is replaced by a symmetry under the action of the operator $C$. Like the permutation operator, $C^2=1$, and $C$ has eigenvalues $\pm 1$. While for $\gamma=0$ we have two underlying XXX models which are fully decoupled, when $\gamma\neq 0$, the two models are coupled, and it might be expected a priori that all degenerate levels split. This is however not the case: the action of $C$ remains reducible, even though the two underlying XXZ models are now coupled. We observe that the spaces in the direct sums all obtain different eigenvalues. For example, considering again the case $L=2$, the eigenvalue with degeneracy  $9$ splits into three eigenvalues  with degeneracies $5$, $3$ and $1$. This is consistent with $U_q(sl(2))$ symmetry. The eigenvalue with degeneracy $6$ however remains six times degenerate for finite $\gamma$: in the corresponding subspace, the action of $C$ is thus reducible, and the two underlying irreducible representations with eigenvalues $C=\pm 1$ remain degenerate.

The case $L=4$ is summarised in table \ref{xxxL4}. The sum of the spaces in the decomposition column is equal to the observed degeneracies of the chain with $L=4$, as written in table \ref{d22degens}. So we observe again that all the direct-sum representations break up for finite $\gamma$, and we conjecture this to be true for arbitrary $L$.


\section{The Bethe Ansatz solution}\label{d22type2}

The advantage of having an open boundary condition that stems from a solution to the boundary Yang-Baxter equation (\ref{reflection}) is that the model should admit an exact solution. In particular, the Bethe Ansatz equations corresponding to the $K$-matrix defined in (\ref{kmat1})--(\ref{kmat2}) have been found in \cite{martinsd22} and \cite{nepomechied22}. When the additive and multiplicative normalisation constants of the Hamiltonian are defined as in (\ref{hopen4}), the Bethe Ansatz solution tells us that the energy eigenvalues are given by
\beq\label{d22eigs}
E_{D_2^2}=\sum\limits_{j=1}^{m}\frac{2\sin^2(2\gamma)}{\cosh2\lambda_j-\cos2\gamma} \,,
\eeq
where the $\lambda_j$ are solutions to the Bethe Ansatz equations (BAE)

\beq\label{k1d22bae}
\left[ \frac{\sinh(\lambda_j+i\gamma)}{\sinh(\lambda_j-i\gamma)} \right]^{2L} = \prod_{k=1,k\neq j}^{m} \frac{\sinh \big(\frac{\lambda_j}{2}-\frac{\lambda_k}{2}+i\gamma \big)}{\sinh \big( \frac{\lambda_j}{2}-\frac{\lambda_k}{2}-i\gamma \big)}\frac{\sinh \big( \frac{\lambda_j}{2}+\frac{\lambda_k}{2}+i\gamma \big)}{\sinh \big( \frac{\lambda_j}{2}+\frac{\lambda_k}{2}-i\gamma \big)} \,.
\eeq
The continuum limit is studied by finite-size scaling of the energy eigenvalues given in (\ref{d22eigs}). We have \cite{Blote1986}
\beq\label{flscaling}
E=f_0L+f_{\rm s}-\frac{\pi v_{\rm F}(\frac{c}{24}-h)}{L}+\mathcal{O}\left( \frac{1}{L^2} \right) \,,
\eeq
where $L$ is the system size, $c$ is the central charge, $h$ is the conformal dimension of the primary field corresponding to the eigenvalue under consideration, $f_0$ is the bulk energy density and $f_{\rm s}$ is the surface energy. The Fermi velocity $v_{\rm F}$ was calculated in \cite{ikhlef2008} and is given by
\beq\label{vfd22}
v_{\rm F}=\frac{2\pi\sin(\pi-2\gamma)}{\pi-2\gamma} \,.
\eeq
It is found that, in the continuum limit, the generating function of levels is
\beq\label{genfunc}
\mathcal{Z}=\sum\limits_{m=0}^{\infty}(2m+1)Z_m \,,
\eeq
where $Z_m$ is the generating function corresponding to the antiferromagnetic Potts model with free boundary conditions, given in \cite{AFPottsSaleur} as
\beq
\label{AFPottsgen}
Z_m=\frac{q^{h_m-\frac{c}{24}}}{\eta^2(q)} \left(1+2 \left[ \sum\limits_{j=1}^{\infty}q^{2m^2+m(2j+1)}-\sum\limits_{j=0}^{\infty}q^{2(m+\frac{1}{2})^2+(m+\frac{1}{2})(2j+1)} \right] \right) \,,
\eeq
where
\beq
h_m =\frac{m(m+1)}{k} \,, \label{critexp}
\eeq
with $m \in \mathbb{Z}$ and $\gamma=\frac{\pi}{k}$. Moreover, $\eta(q)$ is the Dedekind eta function, and $q$ denotes the modular parameter. The central charge $c$ is given by
\beq\label{ccharge}
c=2-\frac{6}{k} \,. 
\eeq
These values for the central charge \eqref{ccharge} and critical exponents \eqref{critexp} will be derived analytically in section \ref{xxzsubset} by mapping some of the solutions to (\ref{k1d22bae}) to solutions of the Bethe Ansatz equations of the open XXZ Hamiltonian with some particular boundary conditions. Section \ref{baesol} will then consider solutions to (\ref{k1d22bae}) that do not correspond to solutions of any XXZ Bethe Ansatz equations. Some examples of these other solutions to (\ref{k1d22bae}) will be presented and the scaling behaviour of the eigenvalues corresponding to these solutions will be shown to reproduce the first few terms in (\ref{AFPottsgen}). In section \ref{TLreps}, the generating function defined in (\ref{AFPottsgen}) will be observed by direct diagonalisation of the Hamiltonian for a range of values of $\gamma$.

\subsection{The XXZ subset}\label{xxzsubset}
\subsubsection{Even number of Bethe roots}\label{evenbroots}
Consider solutions to the BAE (\ref{k1d22bae}) of the form
\begin{subequations}
\label{alphadef}
\begin{eqnarray}
	\lambda_j^0 &=& \alpha_j^0+i\frac{\pi}{2} \,, \\
	\lambda_j^1 &=& \alpha_j^1-i\frac{\pi}{2} \,,
\end{eqnarray}
\end{subequations}
so that (\ref{k1d22bae}) becomes
\beq\label{baesimp1}
\begin{aligned}
\left[ \frac{\cosh(\alpha_j^0+i\gamma)}{\cosh(\alpha_j^0-i\gamma)} \right]^{2L} = &\prod_{k=1,k\neq j}^{\frac{m}{2}} \frac{\sinh(\frac{\alpha_j^0}{2}-\frac{\alpha_k^0}{2}+i\gamma)}{\sinh(\frac{\alpha_j^0}{2}-\frac{\alpha_k^0}{2}-i\gamma)}\frac{\cosh(\frac{\alpha_j^0}{2}+\frac{\alpha_k^0}{2}+i\gamma)}{\cosh(\frac{\alpha_j^0}{2}+\frac{\alpha_k^0}{2}-i\gamma)}\\
&\prod_{k=1}^{\frac{m}{2}} \frac{\cosh(\frac{\alpha_j^0}{2}-\frac{\alpha_k^1}{2}+i\gamma)}{\cosh(\frac{\alpha_j^0}{2}-\frac{\alpha_k^1}{2}-i\gamma)}\frac{\sinh(\frac{\alpha_j^0}{2}+\frac{\alpha_k^1}{2}+i\gamma)}{\sinh(\frac{\alpha_j^0}{2}+\frac{\alpha_k^1}{2}-i\gamma)} \,,
\end{aligned}
\eeq
while the $\alpha_j^1$ can be seen to satisfy a similar equation. Taking the subset of solutions where 
\beq
\label{alphasubset}
 \alpha_k^0=\alpha_k^1 \equiv \alpha_k \,,
\eeq
equation (\ref{baesimp1}) becomes
\beq\label{xxzsubbae}
\left[ \frac{\cosh(\alpha_j+i\gamma)}{\cosh(\alpha_j-i\gamma)} \right]^{2L}\frac{\sinh(\alpha_j-i\gamma)}{\sinh(\alpha_j+i\gamma)} = \prod_{k=1,k\neq j}^{\frac{m}{2}} \frac{\sinh(\alpha_j-\alpha_k+2i\gamma)}{\sinh(\alpha_j-\alpha_k-2i\gamma)}\frac{\sinh(\alpha_j+\alpha_k+2i\gamma)}{\sinh(\alpha_j+\alpha_k-2i\gamma)} \,.
\eeq
Consider now the open XXZ Hamiltonian with boundary fields $H$ and $H'$:
\beq
\mathcal{H}_{\rm XXZ}=-\frac{1}{2}\left[\sum\limits_{i=1}^{L-1}(\sigma_i^x\sigma_{i+1}^x+\sigma_i^y\sigma_{i+1}^y-\cos\gamma_0 \, \sigma_i^z\sigma_{i+1}^z)+H\sigma_1^z+H'\sigma_L^z\right] \,.
\eeq
It was shown in \cite{SALEURBauer} that the eigenvalues of $\mathcal{H}_{\rm XXZ}$ are given by

\beq\label{xxzeigs}
E=-\sum\limits_{k=1}^{m'}\frac{2\sin^2\gamma_0}{\cosh2\mu_k-\cos\gamma_0}+\frac{1}{2}(N-1)\cos\gamma_0+\text{boundary terms} \,.
\eeq
The second term and the boundary terms in (\ref{xxzeigs}) are not important here, since we are interested in looking at the CFT properties in the thermodynamic limit which we can calculate from the terms proportional to $\frac{1}{N}$. The $m'$ Bethe roots $\mu_k$ in (\ref{xxzeigs}) are given by the solutions to the BAE
	\beq\label{xxzbae}
	\left(\frac{\sinh(\mu_j+i\frac{\gamma_0}{2})}{\sinh(\mu_j-i\frac{\gamma_0}{2})}\right)^{2L}\frac{\sinh(\mu_j+i\Lambda)}{\sinh(\mu_j-i\Lambda)}\frac{\sinh(\mu_j+i\Lambda')}{\sinh(\mu_j-i\Lambda')}=\prod\limits_{k\neq j}^{m'}\frac{\sinh(\mu_j-\mu_k+i\gamma_0)}{\sinh(\mu_j-\mu_k-i\gamma_0)}\frac{\sinh(\mu_j+\mu_k+i\gamma_0)}{\sinh(\mu_j+\mu_k-i\gamma_0)} \,,
	\eeq
where the parameters $\Lambda,\Lambda'$ are defined in terms of the boundary parameters $H,H'$ as
\beq
e^{2i\Lambda}=\frac{H-\Delta-e^{i\gamma_0}}{(H-\Delta)e^{i\gamma_0}-1}
\eeq
and similarly for $\Lambda'$. Compare the energies in equations (\ref{xxzeigs}) and (\ref{d22eigs}) and set $\gamma_0=\pi-2\gamma$ as in \eqref{u0gam0}. We then have that
\beq\label{d22eigs2}
E_{D_2^2}=-\sum\limits_{k=1}^{m}\frac{2\sin^2\gamma_0}{\cosh2\alpha_k-\cos\gamma_0} \,,
\eeq
where the $\alpha_k$ were defined in equation (\ref{alphadef}) and subject to \eqref{alphasubset}. Observe that the form of the energy in equation (\ref{d22eigs2}) is precisely the same as the energy of the XXZ chain in equation (\ref{xxzeigs}) if we have $\alpha_k=\mu_k$, up to the boundary and bulk terms that will only contribute to the $\mathcal{O}(1)$ and $\mathcal{O}(N)$ terms which we are not interested in here. We can ensure that $\alpha_k=\mu_k$ by comparing  (\ref{xxzbae}) with (\ref{xxzsubbae}) and setting
\begin{subequations}
\label{lambdadef}
\begin{eqnarray}
 m &=& 2m' \,, \\
\Lambda &=& \frac{\pi}{2}-\frac{\gamma_0}{2} \,, \\
\Lambda' &=& 0 \,,
\end{eqnarray}
\end{subequations}
which ensures that the solutions to (\ref{xxzbae}) with (\ref{xxzsubbae}) are identical and hence
\beq\label{encomp}
E_{D_2^2}=2E_{\rm XXZ} \,.
\eeq

Now we can use the known scaling behaviour of the open XXZ chain to study the scaling behaviour of some states in the \dtt chain,  namely the subset of states satisfying \eqref{alphasubset}. We have from \cite{SALEURBauer} that, for general $\Lambda,\Lambda'$, the effective central charge of the lowest-energy state the XXZ chain (corresponding to the critical exponent $h$) with total magnetisation $S$ is given by
\beq\label{ceff}
c_{\text{eff}}=1-\frac{6}{1-\frac{\gamma_0}{\pi}}\left(1-\frac{\gamma_0+\Lambda+\Lambda'-2\pi S (1-\frac{\gamma_0}{\pi})}{\pi}\right)^2 \,.
\eeq
Using then the fact \cite{Alcaraz1987} that the Fermi veloctiy $v_0$ of the XXZ model is given by $\frac{v_{\rm F}}{2}$ where $v_{\rm F}$ is defined in (\ref{vfd22}),
as well as \eqref{u0gam0}, (\ref{lambdadef}) and (\ref{encomp}),
and setting $\gamma=\frac{\pi}{k}$, we obtain that the effective central charge $\tilde{c}_{\text{eff}}$ of a state in the \dtt model is
\beq
\tilde{c}_{\text{eff}}=2c_{\text{eff}}=2-\frac{6}{k}(1+4S)^2 \,.
\eeq
From the bulk central charge of the staggered six-vertex model \cite{AFPottsSaleur} given in (\ref{ccharge}) and the relationship between the critical exponent $h$ and the effective central charge
\beq\label{hceff}
h=\frac{c-\tilde{c}_{\text{eff}}}{24} \,,
\eeq
we can obtain
\beq
h= -\frac{1}{4k}+\frac{1}{4k}(1+4S)^2=\frac{2S(2S+1)}{k} \,.
\eeq
Setting now $l=2S$ we have:
\beq\label{pottsexp}
h= h_l \equiv \frac{l(l+1)}{k} \,,
\eeq
with $l$ an even integer. The critical exponents of the antiferromagnetic Potts model with free boundary conditions are actually given by (\ref{pottsexp}) for all $l$ integer \cite{Robertson2019}, but the analysis here only recovered the exponents for $l$ even, since we only considered an even number of Bethe roots $m$. Section \ref{oddbroots} will consider solutions to the Bethe Ansatz equations with an odd number of Bethe roots and will recover as well the exponents (\ref{pottsexp}) for $l$ odd.

\subsubsection{Odd number of Bethe roots}\label{oddbroots}
The analysis in section \ref{evenbroots} considered solutions with an even number of Bethe roots and hence recovered the critical exponents of the antiferromagnetic Potts model in equation (\ref{pottsexp}) corresponding to even sectors of magnetisation. We will now consider an odd number of Bethe roots and derive the critical exponents (\ref{pottsexp}) for $l$ odd. Consider solutions to the Bethe Ansatz equations in (\ref{k1d22bae}) of the form in (\ref{alphadef}) but with one additional root, $\lambda_0^0=i\frac{\pi}{2}$. We now have one more root of the form $\lambda_j^0$ than roots of the form $\lambda_j^1$, and this additional root has vanishing real part. We can go through the same analysis that led to (\ref{xxzsubbae}) for the $m$ even case, finding now

\beq\label{xxzsubbaeodd}
\left[ \frac{\cosh(\alpha_j+i\gamma)}{\cosh(\alpha_j-i\gamma)} \right]^{2L}\frac{\sinh(\alpha_j-2i\gamma)}{\sinh(\alpha_j+2i\gamma)} \frac{\sinh(\alpha-i\gamma)}{\sinh(\alpha+i\gamma)} = \prod_{k=1,k\neq j}^{\frac{m-1}{2}} \frac{\sinh(\alpha_j-\alpha_k+2i\gamma)}{\sinh(\alpha_j-\alpha_k-2i\gamma)}\frac{\sinh(\alpha_j+\alpha_k+2i\gamma)}{\sinh(\alpha_j+\alpha_k-2i\gamma)}
\eeq
when $m$ is odd. Now compare the Bethe Ansatz equation in (\ref{xxzsubbaeodd}) to the XXZ Bethe Ansatz equations in (\ref{xxzbae}). When we set
\begin{subequations}
\label{lambdadef2}
\begin{eqnarray}
 m-1 &=& 2m' \,, \\
 \Lambda &=& \frac{\pi}{2}-\frac{\gamma_0}{2} \,, \\
 \Lambda' &=& \pi-\gamma_0 \,,
\end{eqnarray}
\end{subequations}
applying again \eqref{u0gam0}, then the solutions $\alpha_j$ to (\ref{xxzsubbaeodd}) will be the same as the solutions to (\ref{xxzbae}) and we will once again have that the energy of the \dtt chain is equal to twice that of the XXZ chain as in  (\ref{encomp}). Using  (\ref{ceff}) with the $\Lambda, \Lambda'$ taking values in (\ref{lambdadef2}) we find
\beq
\tilde{c}_{\text{eff}}=2c_{\text{eff}}=2-\frac{6}{k}(4S-1)^2 \,.
\eeq
Now using (\ref{hceff}) we finally obtain
\beq
h=\frac{2S(2S-1)}{k} \,,
\eeq
which is equivalent to (\ref{pottsexp}) for $l=2S-1$.

\subsection{Other solutions of Bethe Ansatz equations}\label{baesol}

We have so far managed to use the Bethe Ansatz solution to derive the critical exponents (\ref{pottsexp}) and central charge (\ref{ccharge}) which provides a lot of evidence that the particular boundary conditions under consideration are in the same universality class as the antiferromagnetic Potts model with free boundary conditions. In order to be sure of this, however, we need to check that the full spectrum of the model is consistent with the generating function (\ref{AFPottsgen}). In other words, we have so far only confirmed that the first term in the expansion of $Z_m$ in (\ref{AFPottsgen}) is consistent with the critical exponents (\ref{pottsexp}) derived in sections \ref{evenbroots} and \ref{oddbroots}, but we need to study the excited states in the chain to compare with the other terms. We will do this by finding solutions to the Bethe Ansatz equations (\ref{k1d22bae}) that are not of the form (\ref{alphadef}).

We shall present some solutions for the test case $\gamma=\frac{\pi}{5}$ and show that the results are indeed consistent with (\ref{AFPottsgen}). Section \ref{TLreps} will then show by direct diagonalisation for a range of values of $\gamma$ that (\ref{AFPottsgen}) is indeed the correct generating function of levels for the spin chain. We will consider separately the cases with total magnetisation $n$ equal to two, one and zero in sections \ref{n2}, \ref{n1} and \ref{n0}, respectively. Note that in our notation $m$ is the number of Bethe roots in any given solution to the Bethe Ansatz equations (\ref{k1d22bae}). Solutions with $m=L$ roots correspond to states in the zero magnetisation sector and more generally, when we define:
\beq
m=L-n\, ,
\eeq
the solutions with $m$ Bethe roots correspond to states with magnetisation $n$.

\subsubsection{The $n=2$ sector}\label{n2}
The Bethe Ansatz equations (\ref{k1d22bae}) are more easily handled when cast in logarithmic form:

\beq\label{logbaed22}
2L\log\left( \frac{\sinh(i\gamma+\lambda_j)}{\sinh(i\gamma-\lambda_j)} \right) = 2i\pi I_j + \sum\limits_{k=1,k\neq j}^{m} \left[
\log\left( \frac{\sinh(i\gamma+\frac{1}{2}(\lambda_j-\lambda_k)}{\sinh(i\gamma-\frac{1}{2}(\lambda_j-\lambda_k)}\right)+\log
\left(\frac{\sinh(i\gamma+\frac{1}{2}(\lambda_j+\lambda_k))}{\sinh(i\gamma-\frac{1}{2}(\lambda_j+\lambda_k))}\right) \right] \,,
\eeq
where the $I_j$ are integers introduced as a result of the periodicity of the logarithms. Now consider solutions of the form (\ref{alphadef}). Equations (\ref{logbaed22}) become
\beq\label{logbae2}
\begin{aligned}
&2L\log\left(\frac{\cosh(i\gamma+\alpha_j^0)}{\cosh(i\gamma-\alpha_j^0)}\right)=2i\pi I_j^0+\sum\limits_{k=1,k\neq j}^{m^0}\log\left( \frac{\sinh(i\gamma+\frac{1}{2}(\alpha_j^0-\alpha_k^0)}{\sinh(i\gamma-\frac{1}{2}(\alpha_j^0-\alpha_k^0)}\right)\\
+&\sum\limits_{k=1,k\neq j}^{m^0}\log\left( \frac{\cosh(i\gamma+\frac{1}{2}(\alpha_j^0+\alpha_k^0)}{\cosh(i\gamma-\frac{1}{2}(\alpha_j^0+\alpha_k^0)}\right)+\sum\limits_{k=1}^{m^1}\log\left( \frac{\cosh(i\gamma+\frac{1}{2}(\alpha_j^0-\alpha_k^1)}{\cosh(i\gamma-\frac{1}{2}(\alpha_j^0-\alpha_k^1)}\right)\\
+&\sum\limits_{k=1}^{m^1}\log\left( \frac{\sinh(i\gamma+\frac{1}{2}(\alpha_j^0+\alpha_k^1)}{\sinh(i\gamma-\frac{1}{2}(\alpha_j^0+\alpha_k^1)}\right) \,,
\end{aligned}
\eeq
where $m^0$ and $m^1$ are the number of roots of the form $\lambda_j^0$ and $\lambda_j^1$ respectively. Note that the Bethe numbers $I_j^0$ now take half-integer values when $m^0+m^1$ is even, and integer values when $m^0+m^1$ is odd. An equation similar to (\ref{logbae2}) holds for the $\alpha_j^1$ roots and the Bethe numbers in that case are written as $I_j^1$. It is convenient to define the functions
\begin{subequations}
\begin{eqnarray}
	k(\lambda) &=& -i\log\left(\frac{\cosh(i\gamma+\lambda)}{\cosh(i\gamma-\lambda)}\right) \,, \\
	\theta^0(\lambda) &=& -i\log\left(\frac{\sinh(i\gamma+\frac{\lambda}{2})}{\sinh(i\gamma-\frac{\lambda}{2})}\right) \,, \\
	\theta^1(\lambda) &=& -i\log\left(\frac{\cosh(i\gamma+\frac{\lambda}{2})}{\cosh(i\gamma-\frac{\lambda}{2})}\right) \,.
\end{eqnarray}
\end{subequations}
Equations (\ref{logbae2}) then become

\begin{subequations}
\label{baelogsimp}
\beq \label{baelogsimp1}
\begin{aligned}
2Lk(\alpha_j^0)=2\pi I_j^0&+\theta^0\left(\frac{1}{2}(\alpha_j^0-\alpha_k^0)\right)+\theta^1\left(\frac{1}{2}(\alpha_j^0+\alpha_k^0)\right)\\
&+\theta^1\left(\frac{1}{2}(\alpha_j^0-\alpha_k^1)\right)+\theta^0\left(\frac{1}{2}(\alpha_j^0+\alpha_k^1)\right)
\end{aligned}
\eeq
and
\beq\label{baelogsimp2}
\begin{aligned}
2Lk(\alpha_j^1)=2\pi I_j^1&+\theta^1\left(\frac{1}{2}(\alpha_j^1-\alpha_k^0)\right)+\theta^0\left(\frac{1}{2}(\alpha_j^1+\alpha_k^0)\right)\\
&+\theta^0\left(\frac{1}{2}(\alpha_j^1-\alpha_k^1)\right)+\theta^1\left(\frac{1}{2}(\alpha_j^1+\alpha_k^1)\right) \,.
\end{aligned}
\eeq
\end{subequations}
It is found that the following configuration of Bethe numbers
\begin{subequations}
\label{bethenumbers}
\begin{eqnarray}
I_j^0 &=& j-\frac{1}{2} \,, \\
I_j^1 &=& j-\frac{1}{2}
\end{eqnarray}
\end{subequations}
leads to a unique solution of (\ref{baelogsimp}) corresponding to the lowest-energy state in the particular magnetisation sector under investigation. These are the states that result in the critical exponents (\ref{pottsexp}).

This section will consider the magnetisation sector $n=2$, so that there are a total of $m=L-2$ Bethe roots, and the structure of the Bethe roots corresponding to excited states that are presented here are valid for $\gamma=\frac{\pi}{5}$. To create an excited state in this sector and for this particular value of $\gamma$ we can shift some of the Bethe numbers in (\ref{bethenumbers}) to the right. In particular, if the lowest-energy state with the configuration in (\ref{bethenumbers}) corresponds to a critical exponent $h$, then if we shift the largest $n^0$ Bethe numbers in the set $I_j^0$ by 1, and if we shift the largest $n^1$ Bethe numbers in the set $I_j^1$ by 1, then we will find a solution to the Bethe Ansatz equations in (\ref{baelogsimp1}) and (\ref{baelogsimp2}) that results in a descendent state with critical exponent $h+n^0+n^1$.

\begin{figure}
	\captionsetup{width=0.3\textwidth}
\centering
\begin{minipage}{.5\textwidth}
\includegraphics[width=.9\linewidth]{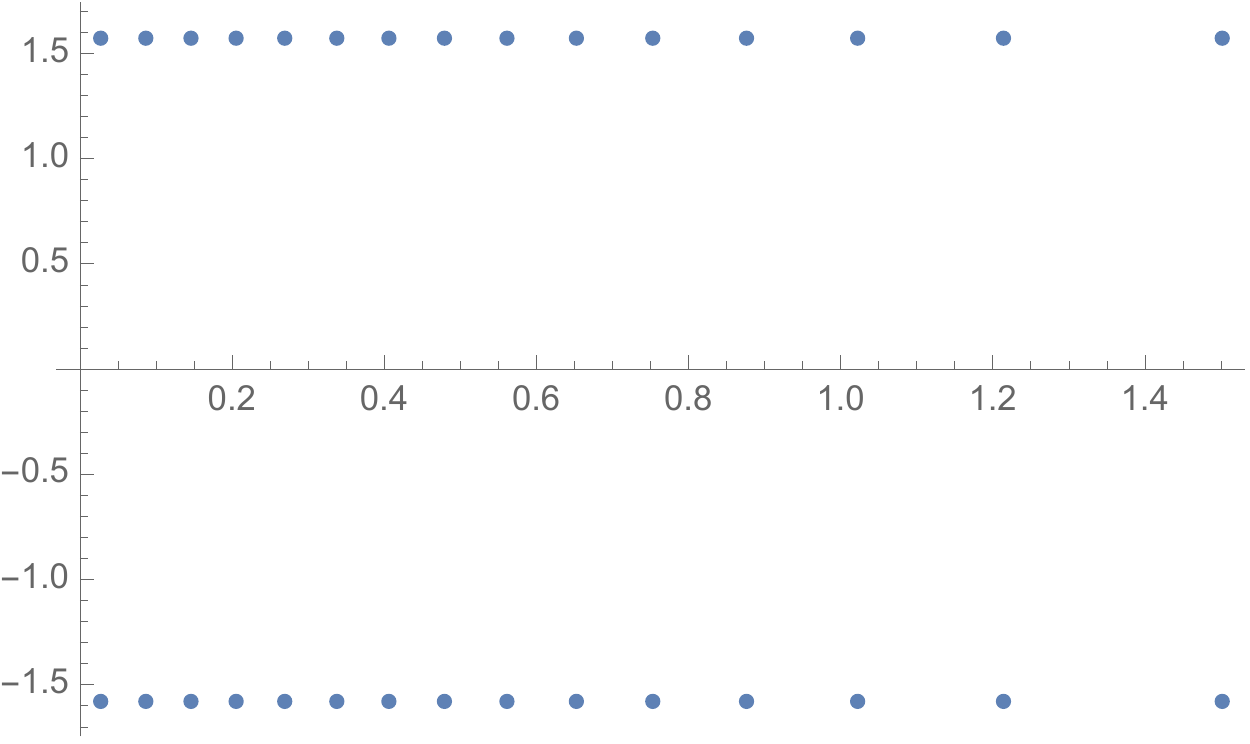}
\captionof{figure}{Bethe roots for $L=32$ corresponding to the lowest-energy state in the $n=2$ sector with $\gamma=\frac{\pi}{5}$.}
\label{groundm2}
\end{minipage}%
\begin{minipage}{.5\textwidth}
  \centering
  \includegraphics[width=.9\linewidth]{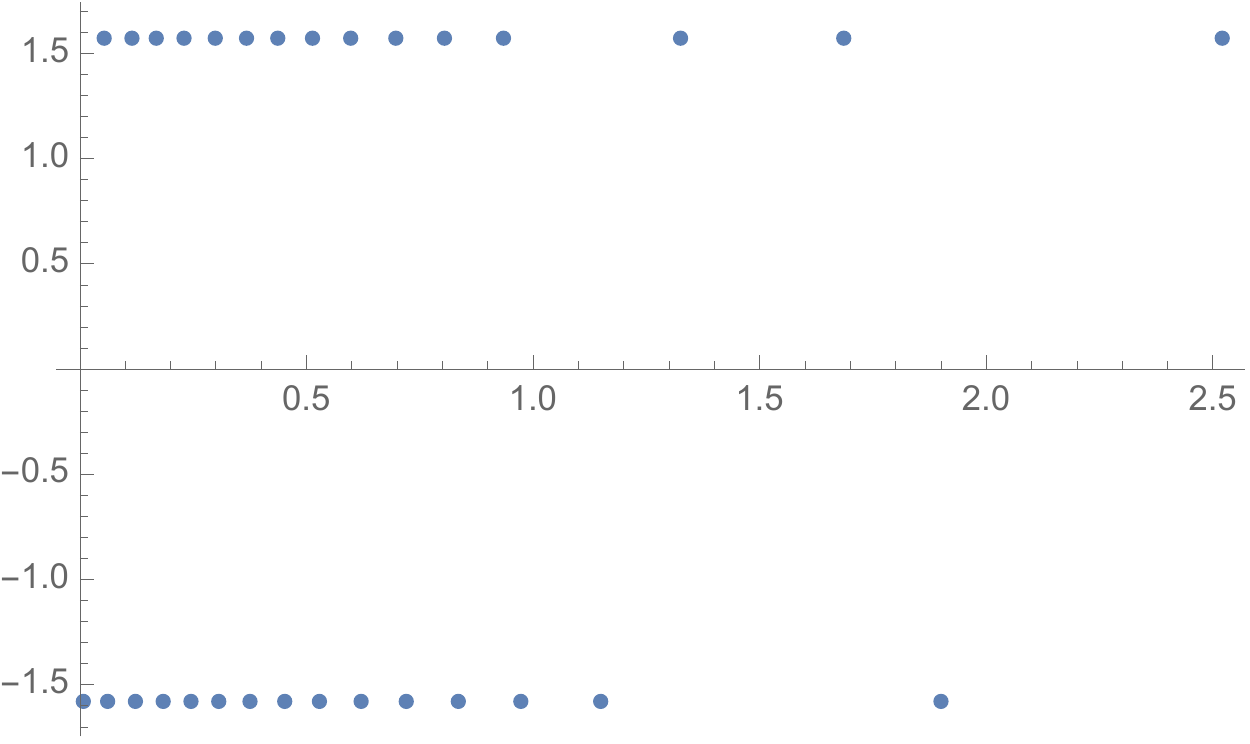}
  \captionof{figure}{Bethe roots for $L=32$ corresponding to an excited state in the $n=2$ sector with $\gamma=\frac{\pi}{5}$. The Bethe numbers on the two lines are shifted by $n^0=3$ and $n^1=1$ respectively.}
  \label{excm2}
  \end{minipage}
\end{figure}

Consider the following example: take $L=8$ and the following configuration of Bethe numbers:
\beq\label{bethenumbersex}
\begin{aligned}
I_1^0&=\frac{1}{2},\ I_2^0&=\frac{5}{2},\ I_3^0&=\frac{7}{2} \,,  \\
I_1^1&=\frac{1}{2},\ I_2^0&=\frac{3}{2},\ I_3^0&=\frac{7}{2} \,.
\end{aligned}
\eeq
These Bethe numbers correspond to a total shift of $n^0+n^1=2+1=3$, hence we expect that in the thermodynamic limit a state of this form corresponds to a critical exponent $h_2+3$ where $h_2$ is the critical exponent in (\ref{pottsexp}) with $l=2$. Observe then that for a given gap of $n^0+n^1$ there are $n^0+n^1+1$ ways to realise this gap, since fixing $n^0+n^1$ there are $n^0+n^1+1$ possible values of $n^0$ which in turn fixes $n^1$.
Examples of solutions of this form are shown in figures \ref{groundm2}--\ref{excm2}.

\begin{figure}
	\captionsetup{width=0.3\textwidth}
\centering
\begin{minipage}{.5\textwidth}
\includegraphics[width=.9\linewidth]{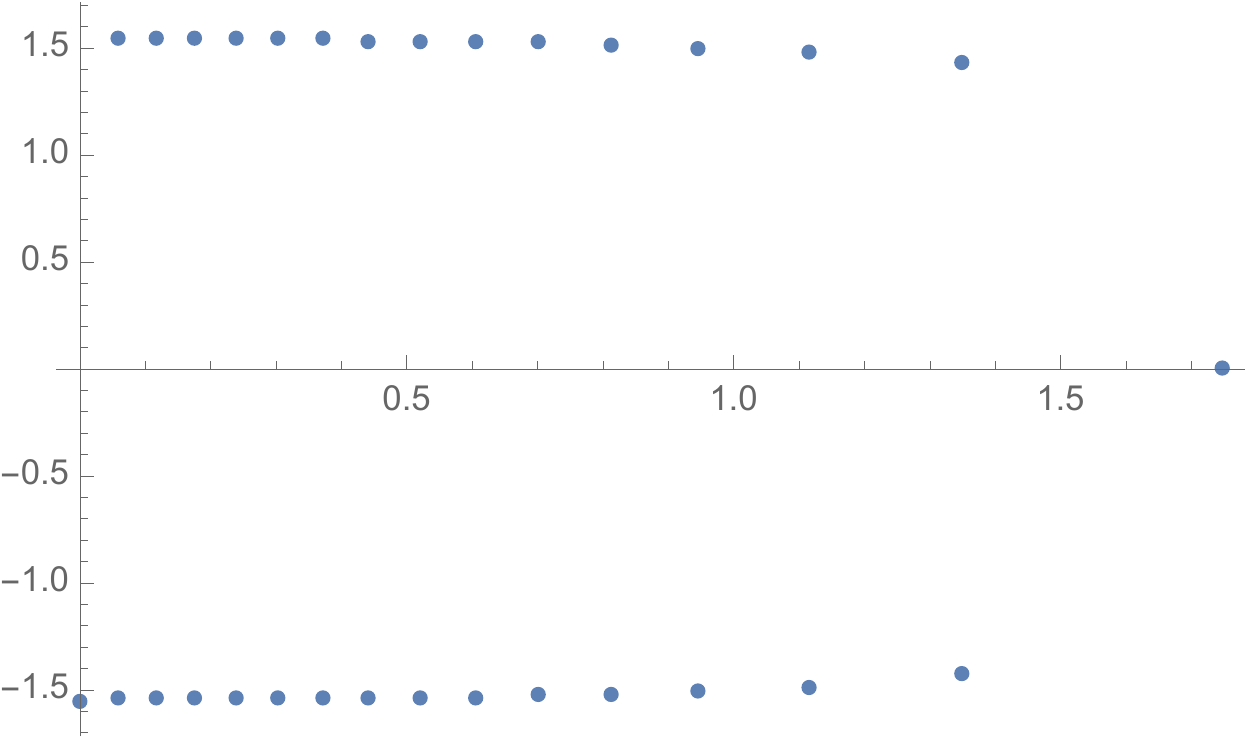}
\captionof{figure}{Bethe roots for $L=32$. A solution to the BAE in the $n=2$ sector with $\gamma=\frac{\pi}{5}$.}
\label{specialm2}
\end{minipage}%
\begin{minipage}{.5\textwidth}
  \centering
  \includegraphics[width=.9\linewidth]{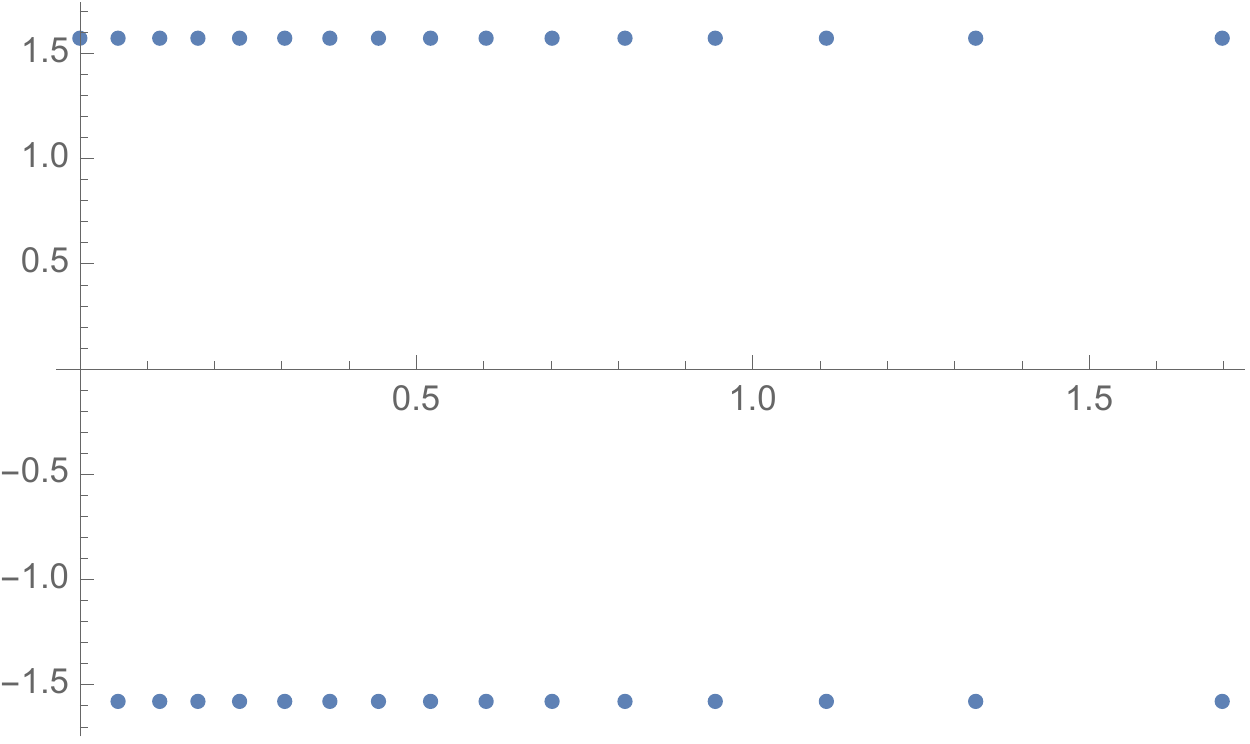}
  \captionof{figure}{Bethe roots for $L=32$. Lowest-energy state in the $n=1$ sector with $\gamma=\frac{\pi}{5}$.}
  \label{groundm1}
  \end{minipage}
\end{figure}

There are solutions to (\ref{k1d22bae}), however, that do not have the form (\ref{alphadef}). An example of a solution of this kind is shown in figure \ref{specialm2}, where there is one root with zero imaginary part, one with zero real part, and all of the other roots have imaginary parts that lie close to $\pm\frac{\pi}{2}$ and are complex conjugates of each other. By studying the scaling behaviour of the state in figure \ref{specialm2} we observe that it corresponds to a critical exponent $h_l+2$ with $h_l$ given by (\ref{pottsexp}) with $l=2$. Using the solutions presented, in addition to the fact that we can always create a new solution to (\ref{k1d22bae}) by shifting all Bethe roots by $+i\pi$, we can reconstruct the first three terms of $Z_2$ in (\ref{AFPottsgen}) for $\gamma=\frac{\pi}{5}$.

\subsubsection{The $n=1$ sector}\label{n1}
\begin{figure}
	\captionsetup{width=0.3\textwidth}
\centering
\begin{minipage}{.5\textwidth}
\includegraphics[width=.9\linewidth]{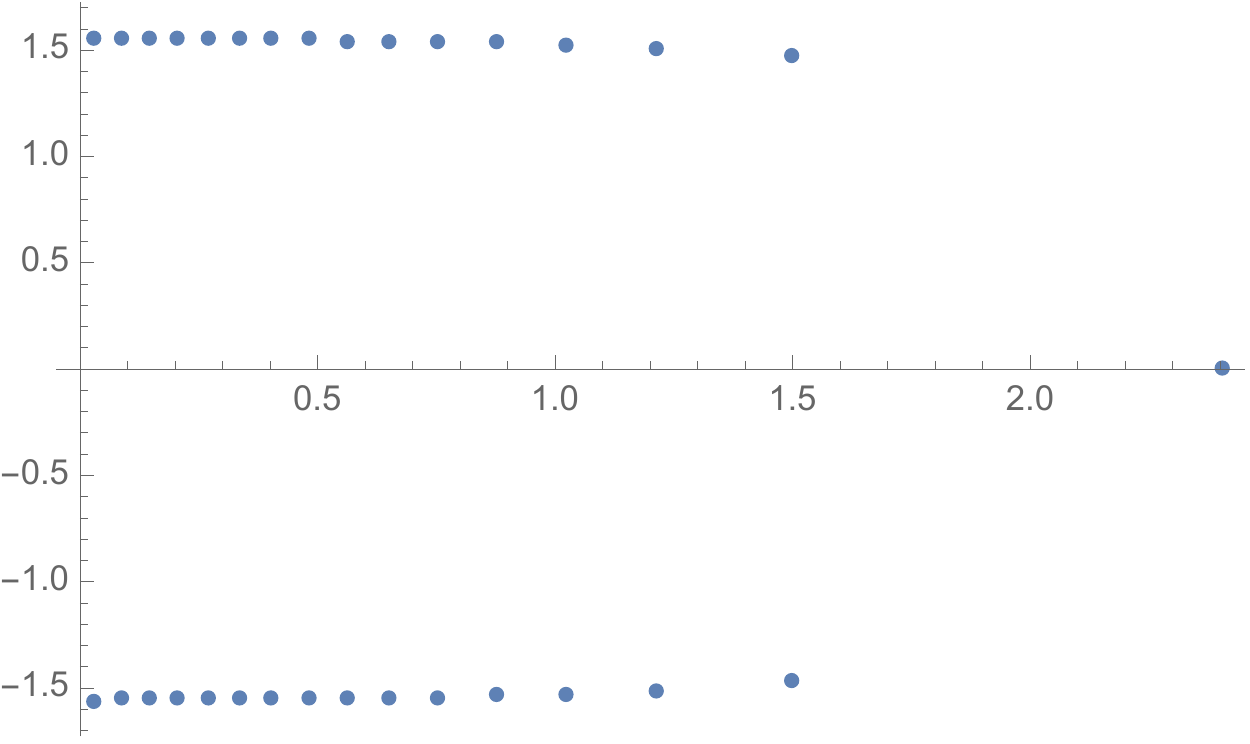}
\captionof{figure}{Bethe roots for $L=32$. First excited state in the $n=1$ sector with $\gamma=\frac{\pi}{5}$.}
\label{excm1}
\end{minipage}%
\begin{minipage}{.5\textwidth}
\includegraphics[width=.9\linewidth]{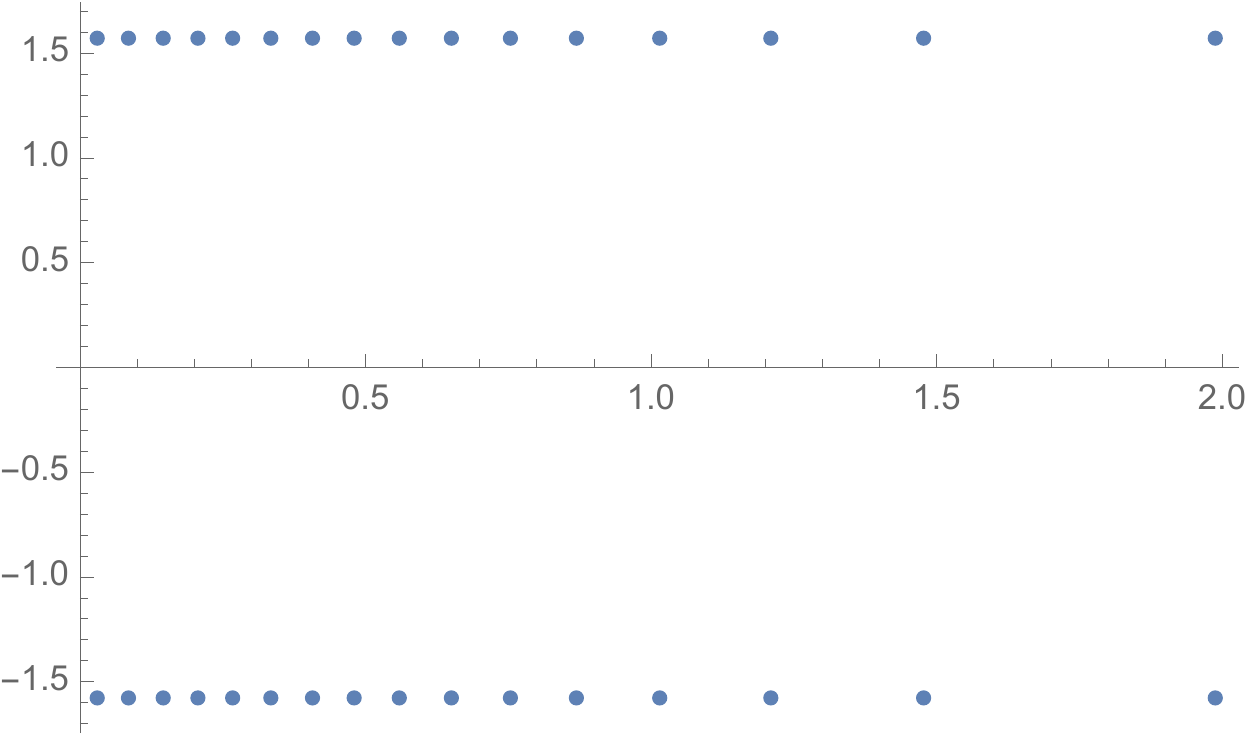}
\captionof{figure}{Bethe roots for $L=32$. Ground state in the $n=0$ sector with $\gamma=\frac{\pi}{5}$.}
\label{groundm0}
\end{minipage}%
\end{figure}

We will now consider an example of solutions to (\ref{k1d22bae}) in the $n=1$ sector, i.e. with $m=L-1$ Bethe roots, again at the particular point $\gamma=\frac{\pi}{5}$. As is the case for all sectors, the solution corresponding to the lowest-energy state is of the form (\ref{alphadef}). An example of such a solution is shown in figure \ref{groundm1}. Since there is an odd number of Bethe roots in the $n=1$ sector, the analysis of section \ref{oddbroots} applies and the critical exponent corresponding to the lowest-energy state is given by (\ref{pottsexp}) with $l=1$. The solution corresponding to the first excited state in this sector is shown in figure (\ref{excm1}). This solution has one Bethe root with zero imaginary part and all of the other roots have imaginary parts that lie close to $\frac{\pi}{2}$ and are complex conjugates of each other. The critical exponent corresponding to this state is given by $h_1+1$ and is therefore consistent with the form of $Z_1$ in (\ref{AFPottsgen}).

\begin{figure}
	\captionsetup{width=0.3\textwidth}
\centering
\begin{minipage}{.5\textwidth}
\includegraphics[width=.9\linewidth]{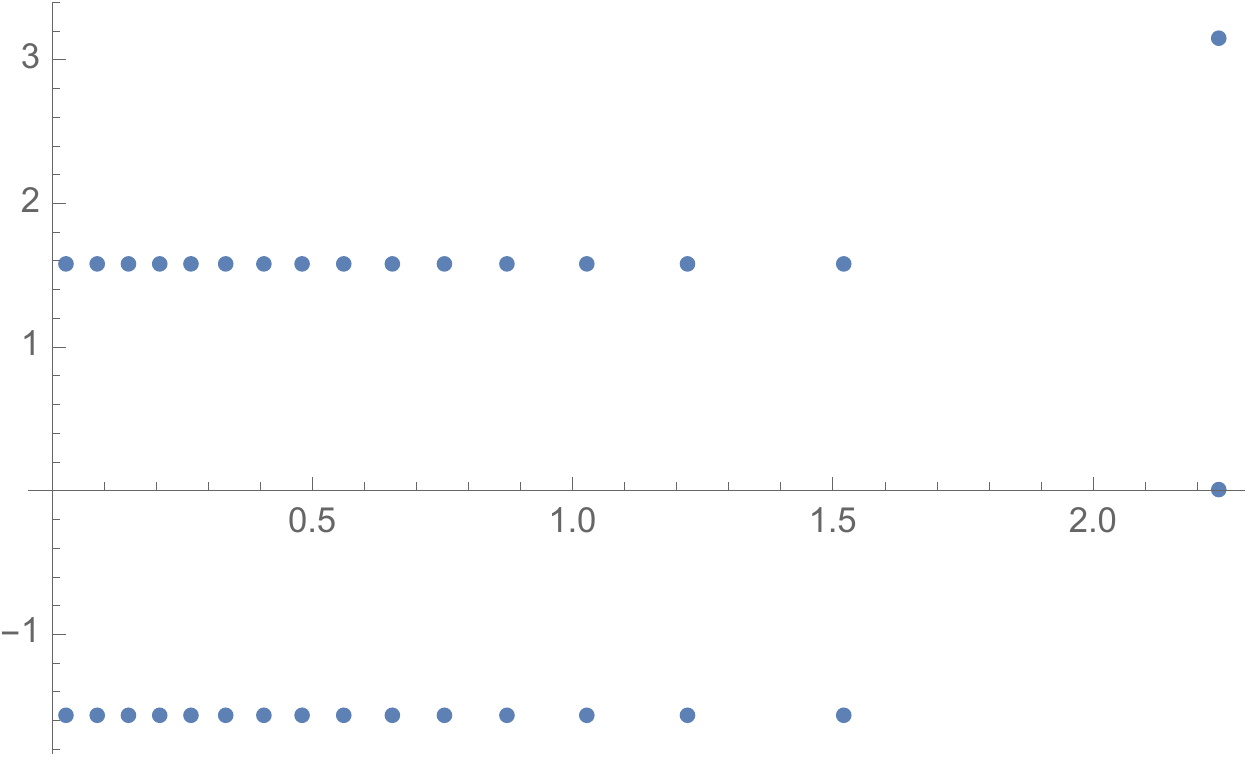}
\captionof{figure}{Bethe roots for $L=32$. First excited state in the $n=0$ sector with $\gamma=\frac{\pi}{5}$.}
\label{excm0}
\end{minipage}%
\begin{minipage}{.5\textwidth}
\includegraphics[width=.9\linewidth]{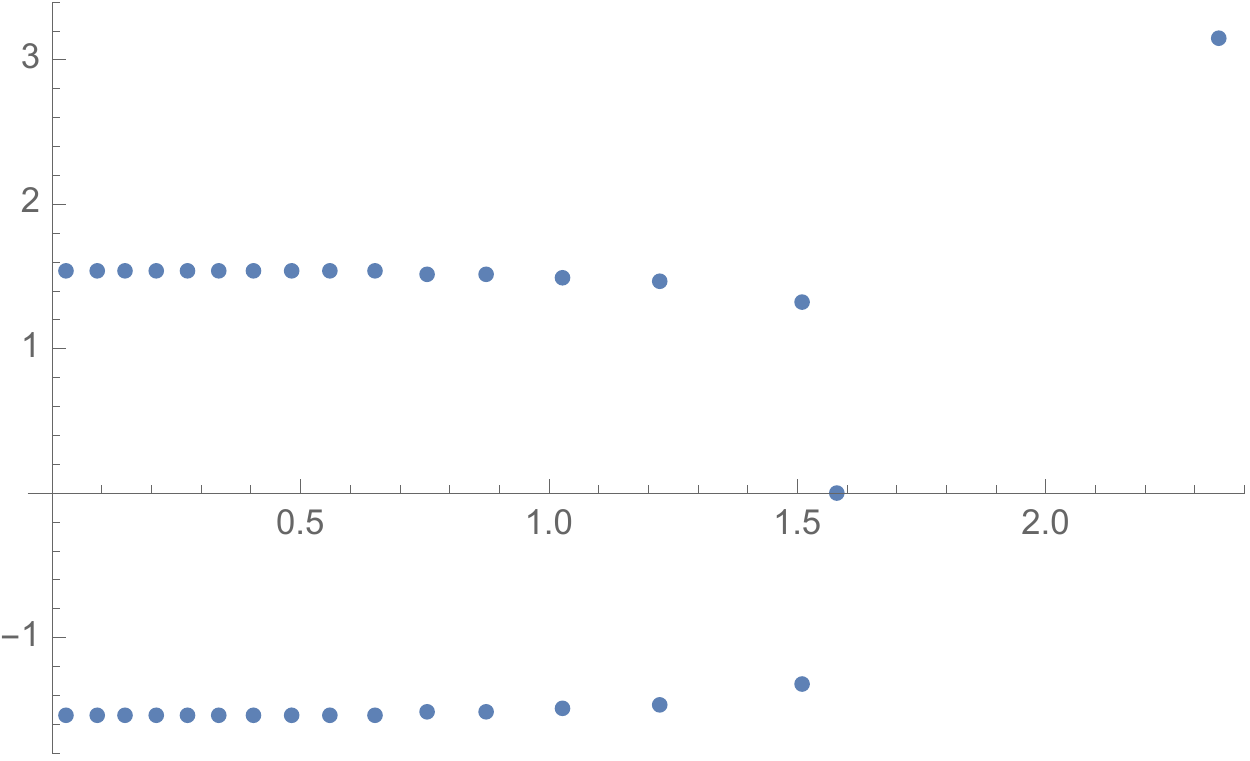}
\captionof{figure}{Bethe roots for $L=32$. Second excited state in the $n=0$ sector with $\gamma=\frac{\pi}{5}$.}
\label{specialm0}
\end{minipage}%
\end{figure}

\subsubsection{The $n=0$ sector}\label{n0}
There are $m=L$ roots in the magnetisation $n=0$ sector. The ground state is of the form (\ref{alphadef}) and a solution for $L=32$ and $\gamma=\frac{\pi}{5}$ is shown in figure \ref{groundm0}. In the thermodynamic limit, a state of this form corresponds to a critical exponent $h=0$, corresponding to $l=0$ in equation (\ref{pottsexp}). The first excited state is of the form shown in figure \ref{excm0}, where we observe that all but two of the roots are on the lines with imaginary part $\frac{\pi}{2}$ and the remaining two roots have imaginary parts $0$ and $\pi$. All of the roots come in pairs differing by $\pm i \pi$. This state results in a critical exponent $h=2$ in the continuum limit, corresponding to the first term of $Z_0$ in (\ref{AFPottsgen}). The next excited state is of the form shown in figure \ref{specialm0}. There is one root with zero imaginary part, one with imaginary part equal to $\pi$, and all of the other roots come in complex conjugate pairs with imaginary parts very close to $\pm \frac{\pi}{2}$. This state results in a critical exponent $h=3$ and corresponds to the next term (\ref{AFPottsgen}). 

\section{Other Temperley-Lieb representations}\label{TLreps}
We have until now been considering the Hamiltonian in (\ref{hopen4}) with the $e_m$ defined in terms of Pauli matrices by (\ref{esigma}). This is known as the vertex-model representation of the TL algebra \eqref{TLrelations}, but there are others representations that we can consider. We will consider the loop representation of the TL algebra in section \ref{looprep}, and the RSOS representation in section \ref{rsosrep}.

\subsection{Loop representation}\label{looprep}
The loop representation of the TL algebra is defined by assigning a graphical representation to each of the $e_i$. Figure \ref{TL} shows diagrams corresponding to this graphical representation of the $e_i$ acting on $N=4$ strands. Multiplication in this representation corresponds to stacking diagrams vertically. The first relation in (\ref{TLrelations}) can then be understood graphically from figure \ref{e1squared} where we see that the formation of a loop is given the Boltzmann weight $\sqrt{Q}$. The graphical form of the second relation is displayed in figure \ref{e1e2e1}. The $e_i$ act on the states in figure \ref{reps}.

We see that in the case $N=4$ the states are divided into three sectors, $\mathcal{W}_0$, $\mathcal{W}_1$ and $\mathcal{W}_2$, where $\mathcal{W}_j$ is the sector with $2j$ through-lines. By definition, a through-line is a connection between the top and the bottom of the diagram. For a system of size $N$ there can be at most $N$ through-lines, and hence the maximum value of $j$ is $\frac{N}{2}$.

We can now study the Hamiltonian in (\ref{hopen4}) with this representation of $e_i$. By directly diagonalising the Hamiltonian it is found that the generating function in the continuum limit, in the sector with $2j$ through lines, is given by $Z_j$, defined in (\ref{AFPottsgen}). The full generating function is then given by
\beq
\mathcal{Z}=\sum\limits_{m=0}^{\infty}Z_m \,,
\eeq
which, when compared to (\ref{genfunc}), is seen to be the same as the generating function in the vertex representation, except that there is a restriction to the highest-weight states of the quantum group symmetry $U_q(sl(2))$.

\begin{figure}
	\centering
\begin{tikzpicture}[scale=0.5]
	
	\node at (-1,1) {$e_1=$};
	
	\draw[black,line width = 1pt] (0,2) .. controls (0.25,1.4) and (0.75,1.4) .. (1,2);
	\draw[black,line width = 1pt] (0,0) .. controls (0.25,0.6) and (0.75,0.6) .. (1,0);
	\draw[black,line width = 1pt] (2,0) -- (2,2);
	\draw[black,line width = 1pt] (3,0) -- (3,2);

	\node at (3.8,0.2) {,};
	
	\node at (5.5,1) {$e_2=$};
	
	\draw[black,line width = 1pt] (7,0) -- (7,2);
	\draw[black,line width = 1pt] (8,0) .. controls (8.25,0.6) and (8.75,0.6) .. (9,0);
	\draw[black,line width = 1pt] (8,2) .. controls (8.25,1.4) and (8.75,1.4) .. (9,2);
	\draw[black,line width = 1pt] (10,0) -- (10,2);
	
	\node at (11.3,0.2) {,};
	
	\node at (13,1) {$e_3=$};
	
	\draw[black,line width = 1pt] (14,0) -- (14,2);
	\draw[black,line width = 1pt] (15,0) -- (15,2);
	\draw[black,line width = 1pt] (16,0) .. controls (16.25,0.6) and (16.75,0.6) .. (17,0);
	\draw[black,line width = 1pt] (16,2) .. controls (16.25,1.4) and (16.75,1.4) .. (17,2);
	
	\node at (18.3,0.2) {,};
	
	\node at (20,1) {$\mathcal{I}=$};
	\draw[black,line width = 1pt] (21,0) -- (21,2);
	\draw[black,line width = 1pt] (22,0) -- (22,2);
	\draw[black,line width = 1pt] (23,0) -- (23,2);
	\draw[black,line width = 1pt] (24,0) -- (24,2);

\end{tikzpicture}
\caption{The graphical interpretation of the Temperley-Lieb loop representation.}\label{TL}
\end{figure}

\begin{figure}
	\centering
\begin{tikzpicture}[scale=0.5]
	
	\node at (-1,2) {$e_1^2=$};
	
	\draw[black,line width = 1pt] (0,2) .. controls (0.25,1.4) and (0.75,1.4) .. (1,2);
	\draw[black,line width = 1pt] (0,0) .. controls (0.25,0.6) and (0.75,0.6) .. (1,0);
	\draw[black,line width = 1pt] (2,0) -- (2,2);
	\draw[black,line width = 1pt] (3,0) -- (3,2);
	
	\draw[black,line width = 1pt] (0,4) .. controls (0.25,3.4) and (0.75,3.4) .. (1,4);
	\draw[black,line width = 1pt] (0,2) .. controls (0.25,2.6) and (0.75,2.6) .. (1,2);
	\draw[black,line width = 1pt] (2,2) -- (2,4);
	\draw[black,line width = 1pt] (3,2) -- (3,4);
	
	\node at (5,2) {$=\sqrt{Q}$};
	
	\draw[black,line width = 1pt] (6.5,3) .. controls (6.75,2.4) and (7.25,2.4) .. (7.5,3);
	\draw[black,line width = 1pt] (6.5,1) .. controls (6.75, 1.6) and (7.25, 1.6) .. (7.5,1);
	\draw[black,line width = 1pt] (8.5,1) -- (8.5,3);
	\draw[black,line width = 1pt] (9.5,1) -- (9.5,3);
	
	\node at (12,2) {$=\sqrt{Q}e_1$};

\end{tikzpicture}
\caption{Graphical interpretation of $e_i^2=\sqrt{Q}e_i$.}\label{e1squared}
\end{figure}

\begin{figure}
	\centering
\begin{tikzpicture}[scale=0.5]
	
	\node at (-2,3) {$e_1e_2e_1=$};
	
	\draw[black,line width = 1pt] (0,2) .. controls (0.25,1.4) and (0.75,1.4) .. (1,2);
	\draw[black,line width = 1pt] (0,0) .. controls (0.25,0.6) and (0.75, 0.6).. (1,0);
	\draw[black,line width = 1pt] (2,0) -- (2,2);
	\draw[black,line width = 1pt] (3,0) -- (3,4);
	
	\draw[black,line width = 1pt] (0,2) -- (0,4);
	\draw[black,line width = 1pt] (1,4) .. controls (1.25,3.4) and (1.75,3.4) .. (2,4);
	\draw[black,line width = 1pt] (1,2) .. controls (1.25,2.6) and (1.75,2.6) .. (2,2);
	
	\draw[black,line width = 1pt] (0,6) .. controls (0.25,5.4) and (0.75,5.4) .. (1,6);
	\draw[black,line width = 1pt] (0,4) .. controls (0.25,4.6) and (0.75,4.6) .. (1,4);
	\draw[black,line width = 1pt] (2,4) -- (2,6);
	\draw[black,line width = 1pt] (3,4) -- (3,6);

	\node at (5,3) {$=$};

	\draw[black,line width = 1pt] (6,4) .. controls (6.25,3.4) and (6.75,3.4) .. (7,4);
	\draw[black,line width = 1pt] (6,2) .. controls (6.25,2.6) and (6.75,2.6) .. (7,2);
	\draw[black,line width = 1pt] (8,2) -- (8,4);
	\draw[black,line width = 1pt] (9,2) -- (9,4);
	
	\node at (11,3) {$=e_1$};
	
\end{tikzpicture}
\caption{Graphical interpretation of $e_1e_2e_1=e_1$.}\label{e1e2e1}
\end{figure}

\begin{figure}
	\centering
\begin{tikzpicture}[scale=0.5]
	
	\node at (-2,0.8) {$\mathcal{W}_0=$};
	
	\node at (-0.5,0.9) {$\{$};
	
	\draw[black,line width = 1pt] (0,1) .. controls (0.25,0.5) and (0.75,0.5) .. (1,1);
	\draw[black,line width = 1pt] (2,1) .. controls (2.25,0.5) and (2.75,0.5) .. (3,1);
	
	\node at (3.5,0.7) {,};
	
	\draw[black,line width = 1pt] (4,1) .. controls (5,0.2) and (6, 0.2) .. (7,1);
	\draw[black,line width = 1pt] (5,1) .. controls (5.25,0.5) and (5.75,0.5) .. (6,1);
	
	\node at (7.4,0.9) {$\}$};

	\node at (-2,-1.2) {$\mathcal{W}_1=$};
	
	\node at (-0.5,-1.1) {$\{$};
	
	\draw[black,line width = 1pt] (0,-1) .. controls (0.25,-1.5) and (0.75,-1.5) .. (1,-1);
	\draw[black,line width = 1pt] (2,-1.5) -- (2,-1);
	\draw[black,line width = 1pt] (3,-1.5) -- (3,-1);
	
	\node at (3.5,-1.4) {,};
	
	\draw[black,line width = 1pt] (4.5,-1.5) -- (4.5,-1);
	\draw[black,line width = 1pt] (5.5,-1) .. controls (5.75,-1.5) and (6.25,-1.5) .. (6.5,-1);
	\draw[black,line width = 1pt] (7.5,-1.5) -- (7.5,-1);
	
	\node at (8,-1.4) {,};
	
	\draw[black,line width = 1pt] (9,-1.5) -- (9,-1);
	\draw[black,line width = 1pt] (10,-1.5) -- (10,-1);
	\draw[black,line width = 1pt] (11,-1) .. controls (11.25,-1.5) and (11.75,-1.5) .. (12,-1);
	
	\node at (12.5,-1.4) {$\}$};

	\node at (-2,-3.2) {$\mathcal{W}_2=$};
	
	\node at (-0.5,-3.1) {$\{$};
	
	\draw[black,line width = 1pt] (0,-3.5) -- (0,-2.8);
	\draw[black,line width = 1pt] (1,-3.5) -- (1,-2.8);
	\draw[black,line width = 1pt] (2,-3.5) -- (2,-2.8);
	\draw[black,line width = 1pt] (3,-3.5) -- (3,-2.8);

	\node at (3.5,-3.1) {$\}$};

\end{tikzpicture}
\caption{The representation spaces of the Temperley-Lieb algebra acting on $N=4$ strands.}\label{reps}
\end{figure}

\subsection{RSOS representation}\label{rsosrep}
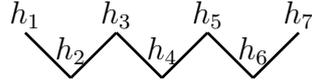
\begin{figure}[ht]
	\centering
\begin{tikzpicture}[scale=1.2]
	
\draw[black,line width = 1pt](1,1)--(1.5,1.5);
\draw[black,line width = 1pt](2,1)--(2.5,1.5);
\draw[black,line width = 1pt](3,1)--(3.5,1.5);
\draw[black,line width = 1pt](0.5,1.5)--(1,1);
\draw[black,line width = 1pt](1.5,1.5)--(2,1);
\draw[black,line width = 1pt](2.5,1.5)--(3,1);

\node at (0.5,1.7) {$h_1$};
\node at (1,1.3) {$h_2$};
\node at (1.5,1.7) {$h_3$};
\node at (2,1.3) {$h_4$};
\node at (2.5,1.7) {$h_5$};
\node at (3,1.3) {$h_6$};
\node at (3.5,1.7) {$h_7$};

\end{tikzpicture}

\caption{The state $\left| h_1h_2 h_3 h_4 h_5 h_6 h_7 \right\rangle$.}\label{RSOSrow}
\end{figure}

In the RSOS representation of the TL algebra, the $e_i$ act on states of neighbouring ``heights'' $h_i=1,2,\ldots,k$, subject to the constraint $|h_{i+1}-h_i| = 1$. We restrict here to $A_k$-type RSOS models. An example of such a state is shown in figure \ref{RSOSrow}. The $e_i$ then take the explicit form \cite{PasquierRSOS}
\beq\label{tlrsos}
e_i \left| h_1,\ldots,h_{i-1},h_i,h_{i+1},\ldots,h_{N+1} \right\rangle =\delta(h_{i-1},h_{i+1})\sum\limits_{h_i'}\frac{\sqrt{S_{h_i}S_{h_i'}}}{S_{h_{i-1}}} \left| h_1,\ldots,h_{i-1},h_i',h_{i+1}, \ldots,h_{N+1} \right\rangle \,,
\eeq
where $\left| h_1,\ldots,h_{i-1},h_i,h_{i+1},\ldots,h_{N+1} \right\rangle$ is the state defined in figure \ref{RSOSrow} and the $S_{h_i}$ are defined as
\beq\label{Sadef}
S_a=\frac{\sin(\frac{a\pi}{k})}{\sin(\frac{\pi}{k})} \,.
\eeq
We recall that $\sqrt{Q}=2\cos(\frac{\pi}{k})$.

It was found in \cite{AFPottsSaleur} that the generating function of the antiferromagnetic Potts model with free boundary conditions in the RSOS representation is given by the string functions $c^0_l$, i.e. the generating function of levels in the $Z_{k-2}$ parafermion CFT. The general form of the string functions $c_l^m$ are given by \cite{Jayaraman}:
\beq
 c_l^m = \frac{1}{\eta(q)^2}\sum_{\substack{n_1, n_2 \in \mathbb{Z}/2  \\ n_1-n_2\in \mathbb{Z} \\ n_1 \geq |n_2|, -n_1 > |n_2| }}(-1)^{2n_1}\text{sign}(n_1)q^{\frac{(l+1+2n_1k)^2}{4k}-\frac{(m+2n_2(k-2))^2}{4(k-2)}} \,,
\eeq
and $l = |h_{N+1} - h_1|$ is the difference between the heights on the left and right boundaries. We observe the same generating function in the continuum limit when we take the RSOS representation of the TL algebra in the Hamiltonian (\ref{hopen4}).

\section{Discussion}
By considering the antiferromagnetic Potts model in its formulation as a staggered six-vertex model we have shown that it is equivalent to the integrable vertex model constructed from the twisted affine \dtt Lie algebra.\\ 

\no Our Bethe Ansatz analysis of the Hamiltonian in (\ref{hopen4}) tells us that it is in the same universality class as the AF Potts model with free boundary conditions, whose transfer matrix studied in \cite{AFPottsSaleur} we do not believe to be solvable by Bethe Ansatz. Since this Hamiltonian is written in terms of Temperley-Lieb generators, we were also able to study its scaling behaviour in the loop and RSOS representations, finding results consistent with our previous observations of the underlying CFT.\\

\no It would be interesting to establish whether the boundary conditions studied here have an analogue in the model for polymers at the theta-point considered in \cite{Vernier2014,Vernier2015}, since there is strong evidence that the latter model has the very same continuum limit as the AF Potts model.
It should be noticed, however, that the polymer model is based on the integrable $A_2^{2}$ chain, different from the \dtt model discussed in the present paper, so the comparison between the boundary critical behaviour of the two models may very well be quite subtle.

\medskip

Returning to the AF Potts model, we are still missing boundary conditions that result in the continuous spectrum established for the corresponding bulk model \cite{ikhlef2008}, something which was also missing from the analysis in \cite{Robertson2019}. 
The study of other integrable boundary conditions \cite{BAE,Nepomechie2D22} in the context of the \dtt model will however lead to the observation of a continuous spectrum in the continuum limit, as we will report in a forthcoming paper.

\subsection*{Acknowledgments}
This work was supported by the European Research Council through the advanced grant NuQFT. The authors would like to thank Rafael I. Nepomechie and Ana L. Retore for helpful discussions.

\end{document}